\documentclass[floats,prb,twocolumn]{revtex4}

\usepackage{graphicx}
\usepackage{docs}

\def\der#1#2{{\partial#1 \over \partial#2}}
\def\be{\begin{equation}}
\def\ee{\end{equation}}
\def\ba#1{\begin{array}{#1}}
\def\ea{\end{array}}
\def\bn{\begin{enumerate}}
\def\en{\end{enumerate}}

\def\rr{\right}
\def\l{\left}

\def\H{\mathcal{H}}
\def\summ{\sum\limits}

\def\G{\Gamma}
\def\ket#1{\l|#1\rr\rangle}
\def\bra#1{\l\langle#1\rr|}
\def\intt{\int\limits}
\def\summ{\sum\limits}
\def\ad#1{\hat{a}^{\dagger}_{#1}}
\def\bd#1{\hat{b}^{\dagger}_{#1}}
\def\a#1{\hat{a}_{#1}}
\def\b#1{\hat{b}_{#1}}
\def\prodd{\prod\limits}
\def\beq{\begin{equation}}
\def\eeq{\end{equation}}
\def\S{\overline{S}}
\begin{document}

\title{Entanglement entropy of the random spin-1 Heisenberg chain}
\author{G. Refael}
\affiliation{Department of Physics, California Institute of
  Technology, MC 114-36, Pasadena, CA 91125}
\author{J.~E.~Moore}
\affiliation{Department of Physics, University of California,
Berkeley, CA 94720}
\affiliation{Materials Sciences Division,
Lawrence Berkeley National Laboratory, Berkeley, CA 94720}
%\pacs{}
%\date{\today}
\begin{abstract}

Random spin chains at quantum critical points exhibit an entanglement  
entropy between a segment of length $L$
and the rest of the chain that scales as $\log_2 L$ with a universal  
coefficient.  Since for {\it pure} quantum critical spin chains this  
coefficient is fixed by the central charge of the associated  
conformal field theory, the universal coefficient in the random
case can be understood as an {\it effective central charge}.
In this paper we calculate the entanglement entropy and
effective central charge of the spin-1 random Heisenberg model in its
random-singlet phase and also at the critical point at which the  
Haldane phase breaks down. The latter is the first entanglement  
calculation for an infinite-randomness fixed point that is {\it not} in the  
random-singlet universality class. Our results are
consistent with a $c$-theorem for flow between infinite-randomness fixed
points. The formalism we use can be generally applied to calculation  
of quantities that depend on the RG history in
$s\ge 1$ random Heisenberg chains.

\end{abstract}

\maketitle

\section{Introduction \label{intro}}

Understanding universal behavior near quantum critical points has been
a major goal of condensed matter physics for at least thirty years.
Quantum critical points describe continuous phase transitions at zero
temperature, where quantum-mechanical phase coherence exists even for
the long-wavelength fluctuations that control the transition.  Some
quantum critical points can be understood via mapping to standard
classical critical points in one higher dimension, but many of the
most experimentally relevant quantum critical points do not seem to
fall into this category.  Furthermore, even quantum critical points
that can be studied using the quantum-to-classical mapping have
important universal features such as frequency-temperature scaling
that do not appear at finite-temperature critical points. \cite{sachdevbook}

One potentially universal feature of quantum critical points is the
ground-state entanglement entropy, defined as the von Neumann entropy
of the reduced density matrix created by a partition of the system
into parts A and B:
\beq
S = - {\rm Tr}\ \rho_A \log_2 \rho_A = - {\rm Tr}\ \rho_B \log_2 \rho_B.
\eeq
The entanglement entropy of the ground state at or near a quantum
critical point can, in some cases, be understood via the
quantum-to-classical mapping; an important example is that of quantum
critical points in one dimension that become two-dimensional conformal
field theories (CFTs), where the entanglement entropy in the quantum
theory has a logarithmic divergence, whose coefficient is connected to the central charge of the classical
CFT:~ \cite{Holzhey94,Vidal03,Calabrese04,Ryu1}
\beq
\lim_{N \rightarrow \infty} S \sim {c \over 3} \log_2 N,
\eeq
where we consider $A$ as a finite contiguous set of $N$ spins and $B$ is the complement of $A$ in the
chain.  Away from criticality, the entanglement $S$ is
bounded above as $N \rightarrow \infty$ (the one-dimensional version
of the ``area law''.\cite{Srednicki93})
Surprisingly, the entanglement entropy, whose definition is closely tied to the
lattice, is actually a universal property of the critical field theory,
and hence independent of lattice details.  At this time we lack a
similarly complete understanding of critical points in higher
dimensions; isolated solvable cases include free
fermions~\cite{gioev,wolf}, higher dimensional conformal field
theories~\cite{Ryu1,Ryu2}, and one class of $z=2$ quantum critical
points.\cite{fradkinmoore}

The connection between the central charge of CFT's and their
entanglement entropy implies that indeed for quantum critical
points with classical analogs, the natural measure of universal
critical entropy in the quantum system (the entanglement entropy) is
determined by the standard measure of critical entropy in the
classical system (the central charge). In addition, it translates
important notions about the central charge to the realm of the
universal quantum measure - the entanglement entropy. Zamolodchikov's
$c$-theorem \cite{Zamolodchikov} states that the central charge $c$
decreases along unitary renormalization-group (RG) flows. Therefore we conclude
that the entanglement entropy of CFT's also decreases along RG
flows. Stated this way, the strength of the $c$-theorem may apply to universal critical entropies in quantum systems that are
not tractable by the quantum-to-classical mapping. 

One such class of systems is the strongly random 1D chains with quantum critical points that can be studied by the
real-space renormalization group (RSRG) technique (see
Ref. \onlinecite{MonthusReview} for review).  The archetype of such a system is the random Heisenberg $s=1/2$
antiferromagnet, which exhibits the ``random-singlet'' (RS) phase.\cite{DSF94} In fact,
disorder introduced to the pure Heisenberg model (which is a CFT with
c=1) is a relevant perturbation, which makes it flow to the RS
phase. In Ref. \onlinecite{RefaelMoore2004} the disorder-averaged entanglement entropy is found
to be logarithmically divergent as in the pure case, but with a
different universal coefficient,
which corresponds to an {\it effective central charge} ${\tilde c}= \ln 2$,
compared to $c=1$ for the pure case.  This result has been
verified numerically for the random-singlet phase of the XX and XXZ
chains, which are expected to have the same critical
properties~\cite{Laflorencie,dechiara}.

The RS phase is but one example of an {\it infinite randomness fixed
point}; special points which obey unique scaling
laws. For instance, instead of energy-length scaling with $1/E\sim
L^z$ for pure quantum-critical scaling, we have: \cite{DSF94,DSF95}  
\be 
\ln 1/E\sim L^{\psi}, 
\ee
where in the RS phase, $\psi=1/2$. The infinite-randomness fixed
points are, loosely speaking, the random analogs of pure CFT's, and
therefore it is important to understand all that we can about their
special universal properties, such as their entanglement
entropy. Generically such points can be reached as instabilities
to disorder of well-known CFT's (e.g. in the XX, Heisenberg, and transverse-field Ising
model). Also, as we shall see, random gapless systems exhibit RG
flow between different infinite randomness fixed points. Another
motivation for the study of the entanglement entropy in these systems
is to understand if they exhibit any correspondence with the pure
CFT $c$-theorem. Namely, does the entanglement entropy, or a related
measure, also decreases along flow lines of systems with randomness?
This question can be broken into two: (a) does the entanglement
entropy decrease along flows between pure CFT's and
infinite-randomness fixed points? (b) does the entanglement entropy
decrease along flows between two different infinite randomness fixed
points?

The first of these questions was taken up by Santachiara,
\cite{Santachiara} who showed that the random singlet phase of the parafermionic Potts model with
$n\ge 42$ flavors has a higher entanglement entropy than the pure
model. The second question, however, is still open. To answer it, and
gain more insight into the entanglement entropy of random systems, we must consider systems which
exhibit more complicated fixed points than the RS phase.
In this paper we analyze entanglement entropy in the fixed points of the random Heisenberg spin-1
chain; particularly at its critical point where the Haldane
phase breaks down (the phase diagram of the spin-1 Heisenberg chain is
shown in Fig. \ref{spin1PD}), we refer to this critical point
throughout as the Haldane-RS critical point. This extends our previous
work by the authors and others on the entanglement entropy in a
variety of  ``random-singlet'' phases \cite{RefaelMoore2004,
Santachiara, YangBonesteel,igloi-2007} and opens the way for a similar calculation in the more
complicated $s>1$ Heisenberg chains.  

The strongly random critical points of $s=1$,~\cite{HymanYang, MGJ,RiegerIgloi,boechat-1996,Continentino-spin-1}
$s=3/2$,~\cite{RKF}, and higher-spin~\cite{DamleHuse} chains are
random analogs of the integrable higher-spin Takhtajan-Babudjian
chains.\cite{takhtajan,babudjian}  Consider $s=1$ as an example: the
Hamiltonian
\beq
H = \sum_i \left[ \hat{\bf S}_i \cdot \hat{\bf S}_{i+1} - (\hat{\bf S}_i \cdot \hat{\bf S}_{i+1})^2\right]
\label{TB}
\eeq
is critical, while without the bi-quadratic term it would be gapped;
its critical theory~\cite{affleckcentral} is referred to as the
$SU(2)$ $k=2$ Wess-Zumino-Witten model, with central charge $c=3/2$.
This critical theory has a relevant operator that corresponds to
modifying the coefficient of the bi-quadratic term in the lattice model
and thereby opening up a gap.  There are similar unstable critical
points in the phase diagram of higher-spin chains: although the
generic higher-spin chain is either gapped (for integer spin) or
gapless (for half-integer spin), there is an integrable Hamiltonian,
given by a polynomial in $(\hat{\bf S}_i \cdot \hat{\bf S}_{i+1})$, that is
gapless and critical with the central charge $c=3s / (s+1)$.  These
critical points have additional symmetry given by the $SU(2)$
Kac-Moody algebra at level $k=2 s$.

Before plunging into the details of our calculation, let us summarize
our results. We find that the leading contribution to the entanglement entropy of the spin-1
random Heisenberg model at the Haldane-RS critical point is:
\be
S\sim\frac{1}{3}{c_{eff}}\log_2 L=\frac{1}{3}\frac{4}{3}\cdot
(1.3327-10^{-3})\cdot \ln 2 \log_2 L.
\ee
where the subtraction indicates the uncertainty in the results, which
is an upper bound. The effective central-charge we find is thus:
\be
c_{eff}^{(r_c)}=1.232
\ee
Referring to question (a) above, $c_{eff}^{r_c}<3/2$, i.e., is smaller
than the central charge of the corresponding pure critical point. With
respect to question (b), this effective central-charge is indeed bigger than the corresponding
central-charge of both the Haldane phase, which vanishes, and the
spin-1 RS phase, which has:
\be
c_{eff}^{s=1\,RS}=\ln 3=1.099.
\ee

In our work we find that the methods used for the
previously studied random-singlet-like critical points are inadequate
to study the spin $s>1/2$ more complicated critical points, where the history
dependence of the renormalization group is more complicated.  The
method developed in this paper and applied to the spin-1 case provides
a general approach to the entanglement entropy of random critical
points in one dimension accessible by real-space renormalization
group. It also presents a well developed framework for other
history dependent quantities, such as correlation functions and
transport properties.

The following section reviews some basic results from the physics of
infinite randomness fixed points, including the higher-symmetry points that
exist in random spin chains.  We review the strategy of our
calculation in Sec. \ref{strategy}, and then proceed to derive the main results of this paper in sections
\ref{history}-\ref{resultsec}, where we obtain the
renormalization-group history dependence for the $s=1$ chain and
thereby its entanglement in a controlled approximation and estimate
the accuracy of the approximation.  The technique is compared to
numerics for a related simplified quantity of ``reduced entanglement'' for which our
technique provides exact results, with good agreement. In Sec. \ref{discussion} we compare our results 
to the known value for a certain fine-tuned
critical point of the pure spin-1 chain, which bears on question (a) above.  We also compare
the entanglement entropy at the two random fixed points of the spin-1
chain.

\section{Review of random-singlet and higher-spin infinite-randomness
  fixed points}

\subsection{Random-singlet phases - the simplest infinite randomness fixed points}

The spin-1/2 Heisenberg chain is an antiferromagnetic chain at
criticality, whose low-energy behavior is described by a
conformal-field theory with central charge $c=1$. Upon introduction of
disorder, the low-energy behavior of the chain flows to a different
critical phase: the random-singlet phase. \cite{DSF94}

The random-singlet phase has very peculiar properties. The Hamiltonian
of the system is:
\be
\H=\summ_i J_i \hat{\bf S}_i\cdot\hat{\bf S}_{i+1}
\ee
Roughly speaking, the strongest bond in the chain, say, $J_i$,
localizes a singlet between sites $i$ and $i+1$. Quantum fluctuations
induce a new term in the Hamiltonian which couples sites $i-1$ and
$i+2$:\cite{MaDas1979,MaDas1980}
\be
\H'_{i-1,\,i+2}=\frac{J_{i-1}J_{i+1}}{2J_i} \hat{\bf S}_{i-1}\cdot\hat{\bf S}_{i+2}.
\label{MaDas}
\ee
Eq. (\ref{MaDas}) is the Ma-Dasgupta rule for the renormalization of
strong bonds. 
Repeating this process produces singlets at all length scales, as
bonds renormalize into large objects. 

A useful parametrization of the couplings in the analysis of the
random spin-1/2 Heisenberg chain is:
\be
\beta_i=\ln\frac{\Omega}{J_i}
\label{betadef}
\ee
where $\Omega$ is the highest energy in the Hamiltonian:
\be
\Omega={\rm max}_i \{J_i\},
\ee
and plays the role of a UV cutoff. It is beneficial to also define a
logarithmic RG flow parameter:
\be
\G=\ln \frac{\Omega_0}{\Omega}
\ee
where $\Omega_0$ is an energy scale of the order of the maximum
$J_i$ in the bare Hamiltonian. In terms of these variables, and using
the Ma-Dasgupta rule, Eq. (\ref{MaDas}), we can construct a flow
equation for the distribution of couplings $\beta_i$:
\be
\ba{c}
\frac{dP(\beta)}{d\G}=\der{P(\beta)}{\beta}\vspace{2mm}\\
+P(0)\intt_0^{\infty}d\beta_1\intt_0^{\infty}d\beta_2
\delta_{\l(\beta-\beta_1-\beta_2\rr)}P(\beta_1)P(\beta_2).
\label{halfflow}
\ea
\ee
Where the first term describes the reduction of $\Omega$, and the
second term is the application the Ma-Dasgupta rule. For the sake of
readability of equations, we denote the convolution with the cross sign:
\be
P(\beta_1)\stackrel{\beta}\times R(\beta_2)=\intt_0^{\infty}d\beta_1\intt_0^{\infty}d\beta_2
\delta_{\l(\beta-\beta_1-\beta_2\rr)}P(\beta_1)R(\beta_2).
\label{crossdef}
\ee

Eq. (\ref{halfflow}) has a simple solution, found by Fisher, which is an attractor to
essentially all initial conditions and distributions: \cite{DSF94}
\be
P(\beta)=\frac{\chi}{\G}e^{-\chi \beta/\G}.
\label{RSdist}
\ee
with $\chi=1$.

Many remarkable features of the random-singlet phase are direct
results of the distribution in Eq. (\ref{RSdist}). In particular, its
energy-length scaling, or, alternatively, the energy scale of a
singlet with length $L$, is:
\be
L^{\psi}\sim \G=\ln 1/E
\label{LEscaling}
\ee
with $\psi$ being a universal critical exponent:
\be
\psi=1/2.
\ee

\subsection{Entanglement entropy in the random-singlet phase}

As mentioned above, in the random singlet phase of the spin-1/2
Heisenberg chain, the entanglement entropy of a segment of length L
with the rest of the chain is: \cite{RefaelMoore2004}
\be
S \sim \frac{\ln 2}{3} \log_2 L. 
\label{oldent}
\ee
The origin of this entanglement in the random singlet phase is simple
to understand. Consider the borders of the segment $L$; the
entanglement of Eq. (\ref{oldent}) is due to singlets connecting the
segment $L$ to the rest of the chain. Each singlet contributes
entanglement 1, and to calculate the total entanglement we need to
count the number of singlets going over one of the borders of the segment
$L$. Alternatively, the entanglement is twice the number of singlets
shorter than $L$ crossing a single partition (i.e., one of the
barriers). 

In Ref. \onlinecite{RefaelMoore2004} we developed a method that allows
to count this number of singlets as the RG progresses from high
energy scales, in which short singlets form, up to the energy scales
$\G\sim L^{1/2}$, where the singlets are of the same length as the
segment length. Our method strongly resembles the calculation in
Ref. \onlinecite{FisherMonthus}. 

In this paper we will generalize this method in order
to calculate the entanglement entropy of more complicated
infinite-randomness fixed points in $s>1/2$ spin-chains. 

The random-singlet phase can occur in the random Heisenberg model of
any spin $s$, but when $s>1/2$, the randomness needs to be sufficiently
strong, such that strong bonds restrict the two sites they connect to be in the complete
singlet subspace. This implies that instead of entanglement 1, each
singlet contributes:
\[
\log_2 (2s+1)
\]
to the total entanglement. Therefore the entanglement of the spin-s
random singlet phase is:
\be
S \sim \frac{\ln 2\cdot \log_2 (2s+1)}{3} \log_2 L. 
\label{oldent-S}
\ee

\subsection{Higher spins infinite-randomness fixed points}

The random-singlet phase, with universal distribution (\ref{RSdist})
and length-energy scaling, Eq. (\ref{LEscaling}), is but one example
of an {\it infinite randomness fixed point}. In general, disordered
systems may have a similar type of logarithmic length-energy scaling
relations, Eq. (\ref{LEscaling}) with different $\psi$. This replaces
the notion of the dynamical scaling exponent $z$ in $L^z\sim 1/E$ of
pure critical points. Similarly, other infinite-randomness fixed
points may exist with a different value of $\chi$ in
Eq. (\ref{RSdist}), but with the same functional form of the
distribution.
The numbers $\psi$ and  $\chi$ are the critical exponents which
parametrize different infinite-randomness universality classes. 

A prime example of different infinite-randomness critical fixed points
comes from Heisenberg models with spin $s>1/2$. Following
Refs. \onlinecite{MGJ,HymanYang} and \onlinecite{RKF} which dealt with the spin-1 and
spin-3/2 cases respectively, Damle and Huse have shown that a spin-s
chain may exhibit infinite-randomness fixed points with:
\be
\ba{cc}
1/\psi=2s+1,\,& \chi=2s.
\ea
\ee
These fixed points were dubbed domain-wall symmetric fixed points for
reasons that will become clear shortly. 

\begin{center}
\begin{figure*}
\includegraphics[width=10cm]{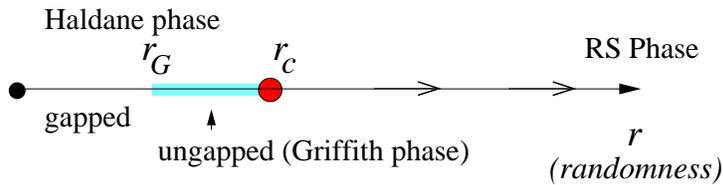}
\caption{Phase diagram of the spin-1 random Heisenberg model. At $r=0$
  the chain is in the gapped Haldane phase and its ground
  state resembles a valence-bond solid (VBS). As randomness is
  increased, the gap is destroyed at $r_G$, but the VBS structure
  survives up to the critical point, $r=r_c$. At $r>r_c$ the chain is
  in the spin-1 random-singlet phase. At the critical point, a
  different infinite randomness fixed point obtains, which has
  $\psi=1/2$, and $\chi=2$. We concentrate on the entanglement entropy
  of this point. \label{spin1PD}}
\end{figure*}
\end{center}

\begin{center}
\begin{figure*}
\includegraphics[width=10cm]{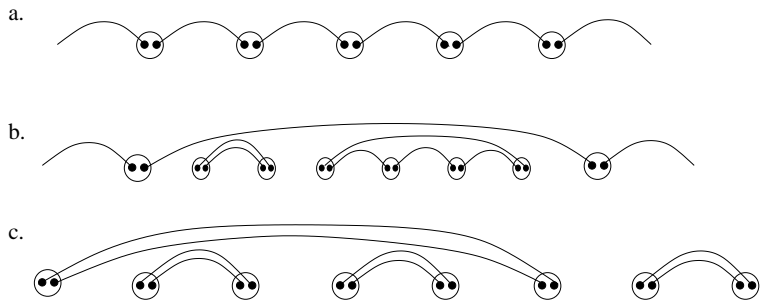}
\caption{(a) At very low disorder, the ground state of the spin-1
  Heisenberg chain is well described as a valence-bond solid. Each
  spin-1 site is described by two spin-1/2 parts (black dots) that are
  symmetrized. Each site forms a spin-1/2 singlet to its right and to
  its left. (b) As disorder grows, defects appear in the VBS
  structure, and the gap is suppressed. (c) At very high disorder, a
  phase transition occurs to the spin-1 random singlet (RS) phase. 
 \label{VBS}}
\end{figure*}
\end{center}

\subsection{The spin-1 Heisenberg model} 

This paper concentrates on the entanglement entropy of the random
spin-1 Heisenberg chain at its critical point. The phase diagram of
this system is shown in Fig. \ref{spin1PD}. Without disorder, the
spin-1 chain has a gap (the ``Haldane gap'').\cite{Haldane} The ground state of the chain
is adiabatically connected to the valence bond solid (VBS) of the AKLT chain, \cite{AKLT} in which each site forms a spin-1 singlet with both of its neighbors: (see Fig. \ref{VBS}a)
\be
\ket{\psi}=\prodd_i
\frac{1}{\sqrt{2}}\l(\ad{i}\bd{i+1}-\bd{i}\ad{i+1}\rr) \ket{0}
\ee
where $\ad{}$ and $\bd{}$ are the creation operators for the spin-up
and spin-down Schwinger bosons. Each spin-1 site can be thought of
consisting of two spin-1/2 parts, symmetrized. In the VBS state it
is those spin-1/2 parts that form the spin-1/2 singlet links over all bonds. These
links are represented by the spin-1/2 singlet creation operator in
terms of the Schwinger bosons: 
\be
\frac{1}{\sqrt{2}}\l(\ad{i}\bd{i+1}-\bd{i}\ad{i+1}\rr).
\label{SBS}
\ee
As disorder increases, defects start occurring in the perfect chain of
singlet links (see Fig. \ref{VBS}b); these defects suppress the gap,
until at a certain disorder $r_G$ the gap vanishes, and the chain enters
into a Griffiths phase.\cite{Damle-griffiths, mdhGriffith} In this region there is still a line of singlets
connecting side to side, but no gap. When a critical disorder $r_c>r_G$
is reached, the connection between the two sides also
disappears, and the Haldane phase terminates. At disorders $r>r_c$ the
chain's ground state is the spin-1 random-singlet state. Our goal is
to calculate the entanglement entropy at $r=r_c$. 

\subsection{Entanglement entropy in the spin-1 Heisenberg model at the
Haldane-RS critical point}

In order to calculate the entanglement
entropy in the spin-1 Heisenberg model at its intermediate-randomness
critical point, we employ the real-space RG technique. This
technique generalizes simply to the spin-$s>1/2$ case. As in the
spin-1/2 case, the first step is to find the strongest bond in the
chain. But instead of putting the strongly interacting sites in a
complete singlet, we just insert one Schwinger-boson singlet (SBS)
(Eq. \ref{SBS}) to the bonds' wave-function. \cite{MGJ} Effectively, this reduces
the spin of each site by 1/2, and reduces the energy of the most
excited state of the bond. If disorder is very weak, the SBS's form
uniformly, and the AKLT state results. On the other hand, if the
disorder is very strong, SBS's form in pairs between strongly
interacting sites, and the spin-1 random singlet phase results. 

Once more, the origin of the entanglement entropy is clear; each SBS
going over a partition contribute entropy {\it of the order} 1
to the entanglement between the two sides of the partition. We need to
count the entanglement of these SBS's until their length equals the
size of the segment under consideration. The challenge of this
calculation, however, is that each singlet going over the partition
contributes an amount of entropy that depends on the singlet configuration forming
both at higher and lower energies. This is due to the fact that each
site can support two singlets connecting it to other sites, and
therefore the ground state of the system is no longer a product state
of site-pairs wave function. 

Another complication in the $s=1$ case is the possibility of
forming ferromagnetic bonds. This happens, for instance, if an even
number of bonds localize one SBS each. The edges of this singlet
chain have spin-1/2 each, which interact with each other
ferromagnetically. At low energies in the RG process, this strong FM
bond can be decimated, in which case the two spin-1/2 coalesce to a
single spin-1. If the partition we are concerned with is between
these two spin-1/2, then after the FM decimation, the partition is
lodged {\it inside} a spin-1 site.  

Thus to be able to calculate the entanglement entropy in $s>1/2$ spin
chains, we need not only a count of the SBS going over a
partition, but a full knowledge of disconnected singlet
configurations: their probabilities, the energy scale at which they
form, and their exact von-Neumann entanglement entropy. To carry out
this calculation, we will use the domain-wall description
\cite{Damle-griffiths,DamleHuse} of the Haldane-RS critical point.  

\subsection{Domain-wall picture}

Our current understanding of the spin-1 critical point between the
Haldane phase and the random-singlet phase relies on the Damle-Huse
domain-wall picture. Let us first demonstrate this picture in terms of
the spin-1/2 random singlet phase. One can think of the random-singlet
phase as forming through a competition between two possible singlet
domains: Domain (1,0) with singlets appearing on odd bonds only, and
domain (0,1) with singlets appearing on even bonds. 
The notation $(a,2s-a)$ with $0\le a\le 2s$ signifies a domain with $a$ spin-1/2
singlet links on odd bonds, and $2s-a$ spin-1/2 singlet links on even
bonds. This notation makes it easy to think about randomness as
competition between different dimerizations. For each
domain, there is a probability $\rho_a$ to be of type $(a,2s-a)$, and
also, for each domain, there is a transfer matrix, which tells the
probability of domain $a$ to be followed by domain $a'$, which is
$W_{aa'}$. Note that:
\[
\summ_{a'=0}^{2s}W_{aa'}=1.
\]

\begin{center}
\begin{figure*}
\includegraphics[width=12cm]{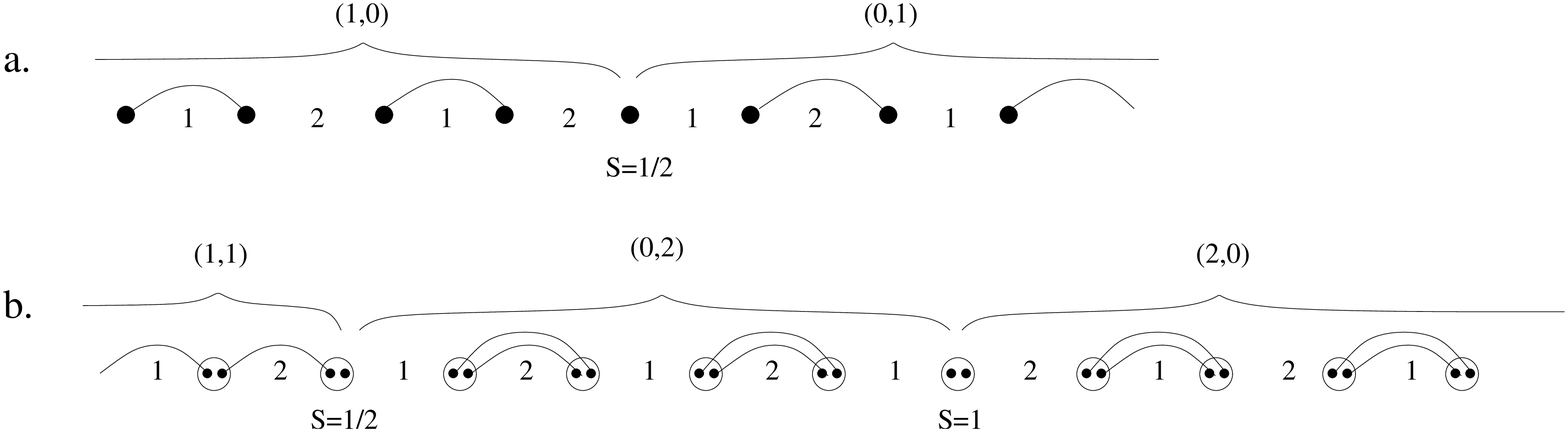}
\caption{(a) Two domains are possible
 in the spin-1/2 Heisenberg model - (1,0) and (0,1). A domain wall between them gives rise to a spin
  1/2 effective site. (b) In the case of a spin-1 chain, there are
  three possible domains: (1,1), (2,0), and (0,2), domain walls
  between them are effectively a spin-1/2 and spin-1 sites respectively. 
\label{domainwall}}
\end{figure*}
\end{center}

In the domain picture, at any finite temperature or energy scale, the
non-frozen degrees of freedom (i.e., spins that were not yet
decimated) lie on domain walls. Thus in the domain wall between the (1,0)
and (0,1) domains, there is one free spin-1/2 site (see
Fig. \ref{domainwall}a). This free spin interacts with similar
spin-1/2's in neighboring domain walls through an interaction mediated
by quantum fluctuations of the domain in between. Thus each domain of
type $a$ is
associated with a bond between neighboring free spins, and has a
distribution of coupling $P_a(\beta)$, with $\beta$ defined in
Eq. (\ref{betadef}). 

The renormalization of strong bonds is now
described as the decimation of a domain. In the spin-1/2 chain,
whenever a domain is decimated its two neighboring domains, being identical, unite to
form a single large domain - a singlet appears over the domain, and
connect the spins on the two domain walls. This is the
Ma-Dasgupta decimation step. The random-singlet phase appears when the
(1,0) and (0,1) domains have the same frequency. It is a
critical point between the two possible dimerized phases associated
with the two domains.

In $s>1/2$ spin chains, the domain picture is richer. In the spin-1
Heisenberg chain there are  three possible domains: (0,2), (1,1), and
(2,0). The VBS is associated with the (1,1) phase, which has a uniform
covering of the chain with spin-1/2 singlet links. On the other hand, the strong
randomness random-singlet phase in this system occurs when the
competing domains are (2,0) and (0,2). This is completely analogous to
the spin-1/2 random singlet phase, except that the domain walls
consist of free spin-1 sites (see Fig. \ref{domainwall}b). 

A general
domain wall between domains $a$ and $a'$ can be easily shown to have
an effective spin:
\be
S_e=\frac{1}{2}|a-a'|
\ee
as each singlet link leaving the domain wall removes a spin-1/2 from
it.

Typically, the decimation of a domain involves forming as many singlet
links as possible between the two domain walls. If the two neighboring
domain are identical, $a'=a''$, then so are the domain walls,
and a full singlet is formed; this is the Ma-Dasgupta decimation rule in Eq. (\ref{MaDas}). If
the two domain-walls are not-identical, and interact with each other
anti-ferromagnetically, singlet links forming between the two domain
walls will exhaust one of the domain-wall spins, and the domain
$D_{a}$ will be swallowed by the domain containing the exhausted spin.

If the two domains neighboring a strong bond are different, $a'\neq
a''$, and the interaction between the two domain-wall spins is
ferromagnetic, the two spins unite into the domain wall between
$D_{a'}$ and $D_{a''}$. For example, consider a (1,1) domain with an even number of links. it
has to connect between a (2,0) domain and a (0,2) domain; both domain
walls will have spin-1/2. Upon
decimation of the (1,1) domain, we are left with a domain wall between
(2,0) and (0,2), which has a spin-1.  

Indeed at the critical point between the Haldane and
random-singlet phases all three domains appear with equal probability - hence
the designation - permutation symmetric critical point. Since each domain appears with the same
frequency at the critical point, each possible domain wall appears with the same
frequency as well. Two domain walls: $(0,2)-(1,1)$ and $(2,0)-(1,1)$ are 
effective spin-1/2's, whereas the third possible domain wall,
$(2,0)-(0,2)$ is a spin-1. Thus, at any finite but low temperature or
energy scale, 2/3 of the unfrozen degrees of freedom are
effectively spin-1/2, and 1/3 are spin-1. These fractions are
universal and a direct consequence of the bare spin of the model. 

The analysis of the domain-theory of the spin-1 critical point is
explained in Ref. \onlinecite{DamleHuse}. The important aspects for
the purpose of our calculations are the following: (1) All domains
have the same bond strength distribution:
\be
P_a(\beta)=\frac{2}{\G} e^{-2\beta/\G},
\label{s11}
\ee
(2) each domain has an equal probability to be of any type:
\be
\rho_a=\frac{1}{2s+1}=\frac{1}{3}
\label{s12}
\ee
(this value indicates the probability after averaging over all
possible neighboring domains), and (3) the transfer matrix $W_{aa'}$
is also the same for all $a\neq a'$:
\be
W_{aa'}=\frac{1}{2s}(1-\delta_{aa'})=\frac{1}{2}(1-\delta_{aa'}).
\label{s13}
\ee

Eqs. (\ref{s11}-\ref{s13}) give a complete description of the spin-1
VBS-to-RS critical point. From it we can deduce the energy-length
scaling properties, Eq. (\ref{LEscaling}):
\be
L\sim \frac{1}{\G^3},
\label{LG}
\ee
i.e.,
\be
L^{1/3}\sim \ln \Omega_0/\Omega.
\ee

\section{Overview of the entropy calculation \label{strategy}}

Combining the above results on this permutation symmetric fixed point,
in the next sections we calculate the entanglement entropy of the
VBS-RS critical point. Unlike the spin-1/2 RS case, where each singlet
contributes exactly entanglement 1, the spin-1 Heisenberg model can
exhibit complicated singlet configuration consisting of overlapping
singlet links, and even of ferromagnetic decimations. 

Nevertheless, the basic principle in our calculation remains similar
to the spin-1/2 calculation outlined in Ref. \onlinecite{RefaelMoore2004}. We can quite generally write the
entanglement entropy of a segment of length $L$ as:
\be
S_{L} \sim \frac{1}{3}c_{eff}\log_2 L=2\frac{\ln \G_L}{\overline{\ell}}\S_{total},
\label{Sintro}
\ee
where, following Eq. (\ref{LG}), $\G_L\sim L^{1/3}$ is the RG flow
parameter at the length scale $L$. Literally, this expression
describes the accumulation of entropy due to configurations with
average entropy $\S_{total}$, which form on average when $\ln\G$
changes by $\overline{\ell}$. This configurations connect the interior
and of the segment to its exterior, on one of its two sides, hence the factor
of $2$. The average entropy $\S_{total}$ is:
\be
\S_{total}=\summ_{c} p_c S_c.
\label{STintro}
\ee
From Eq. (\ref{Sintro}) we can infer the effective central charge:
\be
c_{eff}=\frac{2\S_{total}}{\overline{\ell}} \ln 2
\ee

From Eqs. (\ref{Sintro}) and (\ref{STintro}) the challenge in the
calculation becomes clear. We need to find $\overline{\ell}$, the
probabilities $p_c$ for each configuration, and the entanglement
exhibited in each such configuration, $S_c$.

In Sec. \ref{history} we develop a scheme that calculates the possible configurational
histories in the RG process, and obtains $\overline{\ell}$ as well as
$p_c$ for all configurations. In Sec.\ref{reduced} we define and
calculate exactly a 'reduced entanglement'. This is a simple and rough measure of
entanglement that is determined just by singlet counting. In
Sec. \ref{numerics} we calculate this measure numerically, thus verifying the analytical calculation in Sec. \ref{history}. In
Sec. \ref{conf} we calculate $S_c$ of the various configurations. In Sec. \ref{resultsec} we combine the pieces
into the final answer.

%------------------------------RG   History 

\section{Calculation of the entropy accumulation history dependence \label{history}}

\subsection{Approach to the history dependence}

As pointed out in the introduction, one of the complications in the
spin-1 entropy calculation is the fact that the entanglement of SBS's
depends on the previous and also subsequent RG steps. This arises because SBS
create correlations between partially decimated spins. For this
reason, to find the entanglement entropy we need to calculate the rate of formation of
various finite-size correlated structures. In the following we develop
a system that follows the probability and energy scale of the these
structures. 

Consider a partition across which we calculate the entanglement. Let
us assume that we are at an intermediate stage of the RG, in which
domains are long, and correlations between domain-wall spins can be
neglected (these correlations decay exponentially with the domain's
length). Also, we assume that the partition-bond (the bond in which
the partition is situated) was created by a Ma-Dasgupta decimation
(see Eq. \ref{MaDas}). As we shall see, the last assumption assures
that entanglement from singlets that form in the ensuing RG steps is
independent of the steps preceding the Ma-Dasgupta decimation. 

At this point, when the partition-bond is decimated, there
are three scenarios: 

(1) If the partition-bond is anti-ferromagnetic (AFM), and the
two domains neighboring the bond are different, then one of these domains
must be the (1,1) domain (otherwise we have a (2,0) and
(0,2) domains surrounding a (1,1) domain which must be ferromagnetic
as discussed below). A single
SBS forms over the bond and the partition, and the bond is absorbed to
the (1,1) domain. Only one domain-wall spin is decimated this way, and
the other survives. The entanglement of the singlet just created
depends on what happens next to the undecimated spin. 

(2) If the bond is ferromagnetic (FM), i.e., it corresponds to a (1,1) domain of even
length between a (2,0) and (0,2) domains, the two spin-1/2
domain-wall spins coalesce to form a spin-1 effective site (a domain
wall between the (2,0) and (0,2) domains) which
contains the partition. Needless to say the entanglement depends on what happens to the partition-site in later
stages of the RG. 

(3) If the bond is AFM, but situated between two identical domains,
    the two domain-wall spins connected by the bond are completely
    decimated. The entanglement entropy of this event can only depend
    on the previous decimation history of these domain walls. This is
    the case of the Ma-Dasgupta decimation. Indeed this decimation
    directly follows a Ma-Dasgupta decimation as assumed above, it
    contributes $S_E=1$ between two spin-1/2's or $S_E=\log_2 3$
    between two spin-1's. 

Whereas the third scenario returns the partition-bond to the starting
point of the discussion, the entropy of cases (1) and (2) above will
be determined by the ensuing decimation of the sites near or at the
partition until a Ma-Dasgupta decimation occurs. This is a general principle for all spin-S models;
once the partition-bond is freshly determined by a Ma-Dasgupta
decimation, it means that {\it all} spin-excitations within the domain
are gapped out, and quantum fluctuations above the gap give the
suppressed Ma-Dasgupta coupling. To count the entanglement entropy we
need to follow the decimation history of the partition-bond starting
right after a Ma-Dasgupta decimation, until the next one. 

Our calculation will follow the above lines; we will count how long it
takes, in terms of the RG progression, to form various configurations
between two Ma-Dasgupta decimations. Since there are three possible
domains in the spin-1 Heisenberg chain, we need to consider each
possible domain separately. In Ref. \onlinecite{DamleHuse}, Damle and Huse show that at the spin-1
critical point, each of the three possible domains,
$(2,0),\,(0,2),\,(1,1)$, appear with the same probability. Thus, right
after a Ma-Dasgupta decimation step, the bond over our partition, has
probability $1/3$ of being each of the domains. Due to right-left reflection
symmetry, the contributions due to the bond being domains (2,0) and
(0,2) are the same. Therefore we only need to consider two
possibilities. 

The main complication in our calculation is the possibility of
ferromagnetic decimation steps. A FM decimation renders our partition lodged
inside a spin-1 site, which is also a domain wall. Our
calculation is simplified by splitting the history analysis of the
partition-bond to events from a Ma-Dasgupta decimation up to the
formation of a FM partition-site, and events following the
partition-site formation until a Ma-Dasgupta decimation. 
We start by analyzing the latter. 

\subsection{Note on additivity and independence of RG times} 

One of our goals is the calculation of the energy scales it takes
uncorrelated configuration to form. As shown in
Ref. \onlinecite{RefaelMoore2004}, probability functions of
configurations should be calculated as a function of 
\be
\ell=\ln \G/\G_0.
\label{ellG}
\ee 

In the next sections we will use the fact that this time $\ell$ is
additive in the following sense: the RG time between the formation of a partition-site
(i.e., a spin-1 site containing the partition) to a Ma-Dasgupta
decimation into a partition-bond is independent of the RG history
before the formation of the spin-1 partition-site. This is clear since
all the information about the coupling strengths of the bonds that led
to the formation of the partition-site is swallowed within the
composite spin-1 site, and the ensuing decimations are only determined by bonds
coupling the partition-site to the rest of the chain.  
Thus the RG-time it takes to go between Ma-Dasgupta decimations is the
sum of the time for a partition-site to form, and for the
partition-site to be decimated:
\be
\ell_{MD-MD}=\ell_{MD-PS}+\ell_{PS-MD}
\label{add}
\ee

\begin{center}
\begin{figure*}
\includegraphics[width=15cm]{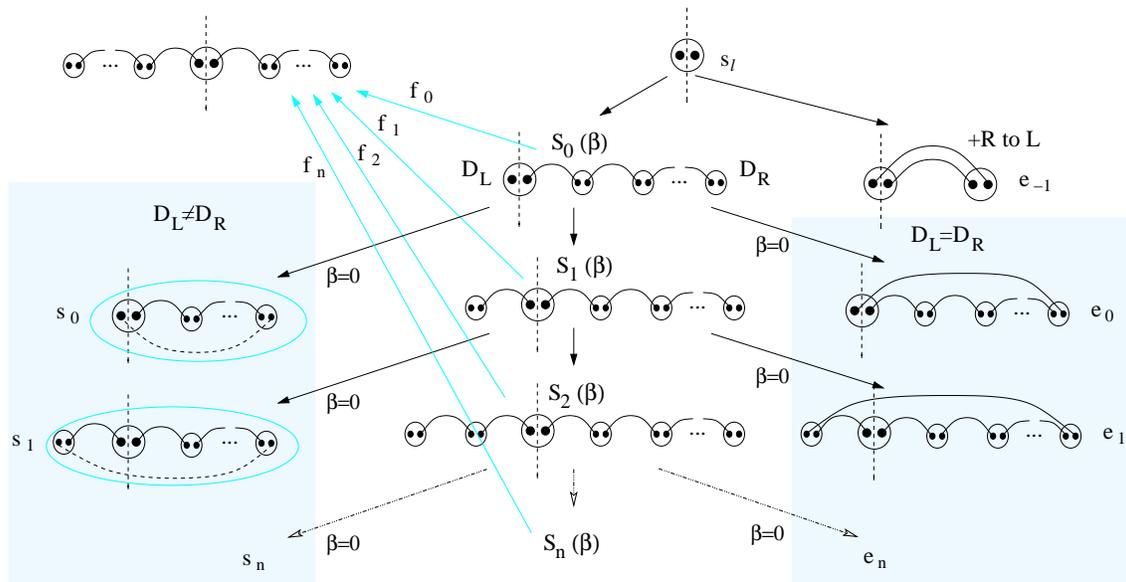}
\caption{The possible outcomes of a spin-1 site when our partition is
  located inside the site. When the bond to the right (left) of the
  site gets decimated, it either forms a complete singlet - in case
  the domain to the left of the bond is the same as that to its
  right - or makes the spin join a (1,1) domain, in which case the
  bond strength has probability distribution $S_0(\beta)$. These two possibilities
  occur with the same probability, 1/2, due to the domain-wall
  permutation symmetry. The (1,1) domain can be decimated by forming
  another singlet (eventuality $e_0$), by forming a triplet
  (eventuality $s_0$) or by joining on its left (right) another (1,1)
  domain through a Ma-Dasgupta decimation (eventuality $f_0$). It could also form another
  bond to its left, giving rise to the next line of diagrams, and the
  eventualities $S_1(\beta),\,e_1,\,s_1,\,f_1$. The subscript
  indicates the number of bonds to the left of the partition  that are
  involved in the (1,1) domain. Note that the $e_n$ and $s_n$
  eventualities occur on a partition-bond decimation ($\beta=0$), when $D_L=D_R$ and $D_L\neq D_R$ respectively,
  and thus with the same probability due to domain-wall permutation
  symmetry ($D_{R/L}$ are the domains to right/left of the bond). 
\label{figRG1}}
\end{figure*}
\end{center}

\subsection{Starting with a spin-1 site \label{spin1}}

Let us now follow the decimation process from the point that the
partition is lodged inside a spin-1. We denote $s(\ell)$ as the
probability of the partition to be inside an untouched spin-1
site. Due to the additivity principal, Eq. (\ref{add}), we can use the
boundary condition:
\be
s(\ell=0)=1
\ee
and follow how this probability decays as the partition-site becomes
correlated through SBS's to its environment. 

The possible RG histories of a spin-1 cluster with distinct entropy
formation are shown in
Fig. \ref{figRG1}. Each of the eventualities in the figure has a
probability distribution $S_n(\beta),\,e_n,\,s_n,\,f_n$, as a function of
$\ell=\ln \G$. Note that in the $e_n$ and $f_n$
eventualities the Ma-Dasgupta rule is applied and we get a new bond
which contains the partition; at this point the spins we followed are
singlet-ed out, and the entanglement of the configuration is unaffected
by the subsequent RG stages. The $s_n$ eventualities return us to the original state of
this part of the calculation, i.e., a partition contained within a
site. The need to separate these eventualities by the subscript $n$,
of the number of effective sites to the left (right) of the partition
is necessary since configurations with different $n$'s produce
different entanglement entropies. 

Given the scaling form of the bond-strength
distribution, 
\be
P(\beta)=\frac{2}{\Gamma}\exp(-2\beta/\G),
\ee
we can now calculate exactly the probability evolution of all eventualities. First, the probability
$s_{\ell}$ will diminish due to decimations of bonds on either of its
sides:
\be
\frac{ds(\ell)}{d\G}=-2P(0)s(\ell).
\label{elleq}
\ee
and recall that $\ell$ and $\G$ are related by Eq. (\ref{ellG}).

Starting with $e_{-1}$, the immediate-singlet eventuality:
\be
\frac{de_{-1}(\ell)}{d\G}=\frac{1}{2}s_{\ell}P(0).
\ee
The interpretation of this equation is that the change in probability
of the $e_{-1}$ eventuality, is $P(0)d\G$ - the probability that the
bond to the right (left) is decimated, times the probability that the
domains to the left and right of the bond are the same, which is
$1/2$, times the probability of the initial configuration,
$s_{\ell}$. Upon substituting $P(0)$ we obtain:
\be
\frac{de_{-1}(\ell)}{d\ell}=s_{\ell}.
\ee

If the bond to the right (left) of the partition connects to a (1,1)
domain, then the spin joins a long (1,1) domain to its right
(left). This domain has coupling strength $\beta$ between its walls,
and hence this eventuality is characterized by the distribution:
$S_0(\beta)$. This distribution can evolve under the RG due to
decimations of other bonds to the right (left) of the partition, where
the (1,1) domain is. But any decimation of its neighboring bonds will
remove probability from $S_0(\beta)$, as will a Ma-Dasgupta decimation of the
bond on which the partition sits, in the case of $\beta=0$. Thus the
flow equation for $S_0(\beta)$ is:
\begin{widetext}
\be
\frac{dS_0(\beta)}{d\G}=\der{S_0(\beta)}{\beta}+\frac{1}{2}P(0)(S_0(\beta)+P(\beta_1)\stackrel{\beta}\times
S_0(\beta_2))-2P(0) S_0(\beta)+\frac{1}{2}s_{\ell}P(0) P(\beta).
\label{S0eq}
\ee
\end{widetext}
The last term represents what happens when the bond to the right
(left) of the partition site is decimated, and the second-nearest domain is (1,1), which
happens with probability $1/2$, hence the factor
of $1/2$. The second-to-last term describes the removal of probability
due to decimations on either side of the partition-bond after it
was formed. The brackets describe what happens when the bond to
the right (left) of the partition-bond is decimated.  The two factors of $1/2$ in the brackets are the probability of the second-nearest domain on
the right (left) being different than (1,1) in the first term, and
(1,1) in the second term. The first term on the RHS describes the
reduction of the energy scale $\Omega$. The cross denotes convolution,
as defined in Eq. (\ref{crossdef}).

But a decimation to the left (right) of the
partition can have two outcomes - if the bond to the left (right) is
connecting the (1,1) domain of the partition with another (1,1) domain
to its left (right), then a decimation of this bond means applying the
Ma-Dasgupta decimation, and we can stop following the RG; this is
$f_0$ outcome, and its evolution is:
\be
\frac{df_{0}(\ell)}{d\G}=\frac{1}{2}P(0)\intt_0^{\infty} d\beta S_0(\beta).
\label{f0eq}
\ee
Alternatively, if the second-nearest domain to the left (right) is not
(1,1), then a decimation of the bond to the left will create another
spin-1/2 singlet over it, and will lead to the distribution
$S_1(\beta)$. $S_1(\beta)$ has the same structure of the flow
equation, except for the source term, which is now fed from $S_0$,
rather than $s_{\ell}$:
\begin{widetext}
\be
\frac{dS_1(\beta)}{d\G}=\der{S_1(\beta)}{\beta}+\frac{1}{2}P(0)(S_1(\beta)+P(\beta_1)\stackrel{\beta}\times
S_1(\beta_2))-2P(0) S_1(\beta)+\frac{1}{2}P(0) S_0(\beta).
\label{S1eq}
\ee
\end{widetext}
again the factor of $1/2$ in the source term is due to the probability
of the second-nearest domain on the left not being (1,1). 

Going back to $S_0(\beta)$, when the bond on the partition is strong,
it gets decimated. If the two neighboring domains are the same, then we apply the Ma-Dasgupta rule once $\beta=0$, in
which case we obtain the singlet-ed cluster $e_0$. The evolution of
this probability is:
\be
\frac{de_{0}(\ell)}{d\G}=\frac{1}{2} S_0(0).
\label{e0eq}
\ee
where the factor of $1/2$ is the probability that the two neighboring
domains are the same. If they are different, a new spin-1 site is
born, which is eventuality $s_0$. It is easy to see that:
\be
\frac{ds_0}{d\G}=\frac{de_0}{d\G}.
\label{s0eq}
\ee
Note that we ignore what happens to the resulting new spin-1
configurations; subsequent RG of this configurations are the same as
those considered in this section, and can be analyzed self-consistently.

Eq. (\ref{s0eq}) completes the consideration of all first-level eventualities. 
It is easy to see that Eqs. (\ref{f0eq}) - (\ref{s0eq}) generalize to
the case of starting with $S_n(\beta)$, which is the distribution of a
bond with $n$ singlets to its left (right), and a long (1,1) domain to
its right (left). Thus the equations for $f_n,\,e_n,\,s_n$ become:
\begin{widetext}
\be
\ba{c}
\frac{df_{n}(\ell)}{d\G}=\frac{1}{2}P(0)\intt_0^{\infty} d\beta
S_n(\beta),\vspace{2mm}\\
\frac{dS_n(\beta)}{d\G}=\der{S_n(\beta)}{\beta}+\frac{1}{2}P(0)(S_n(\beta)+P(\beta_1)\stackrel{\beta}\times
S_n(\beta_2))-2P(0) S_n(\beta)+\frac{1}{2}P(0) S_{n-1}(\beta),\vspace{2mm}\\
\frac{de_{n}(\ell)}{d\G}=\frac{1}{2} S_n(0),\vspace{2mm}\\
\frac{ds_n}{d\G}=\frac{de_n}{d\G}.
\ea
\label{neq}
\ee
\end{widetext}

Before solving all equations, we note that the hardest part of solving Eqs. (\ref{f0eq})-(\ref{neq}) is finding
the distributions $S_n(\beta)$. Here, an important simplification occurs:
Eqs. (\ref{S0eq}) and
(\ref{neq}) for $S_n$ always admit a solution of the form:
\be
S_n(\beta)=\alpha_n P(\beta).
\label{alphaville}
\ee
Substituting Eq. (\ref{alphaville}) in the equation for $S_n$ gives:
\be
\ba{c}
\frac{d\alpha_0}{d\G}=-4\frac{\alpha_0}{\G}+\frac{1}{2}P(0) s_{\ell},\vspace{2mm}\\
\frac{d\alpha_n}{d\G}=-4\frac{\alpha_n}{\G}+\frac{1}{2}P(0).
\alpha_{n-1}
\label{alpha2}
\ea
\ee
Note that $\intt_0^{\infty} d\beta S_n(\beta)=\alpha_n$. 

Starting with Eq. (\ref{elleq}) it is easy to see that:
\be
\G\frac{ds_{\ell}}{d\G}=\frac{ds_{\ell}}{d\ell}=-4 s_{\ell}
\ee
and:
\be
s_{\ell}=e^{-4\ell}.
\ee
$e_{-1}$ is immediately obtained as:
\be
e_{-1}=\frac{1}{4}\l(1-e^{-4\ell}\rr)
\ee
Note that this is the probability for a double singlet forming to the
left, and the same probability applies to singlet-formation to the right. Hence the total probability of a double
singlet forming is $1/2$.

Next is $\alpha_0$:
\be
e^{4\ell}\alpha_0=\intt_0^{\ell} e^{4\ell} s_{\ell}
\ee
In fact, Eq. (\ref{alpha2}) is:
\be
e^{4\ell}\alpha_n=\intt_0^{\ell} d\ell e^{4\ell} \alpha_{n-1}
\ee
It is thus helpful to define:
\be
\tilde{\alpha}_n=e^{4\ell}\alpha_n.
\label{talpha}
\ee
In terms of the $\tilde{\alpha}$ we have:
\be
\ba{c}
\tilde{\alpha}_0=\intt_0^{\ell}d\ell 1=\ell\vspace{2mm}\\
\tilde{\alpha}_n=\intt_0^{\ell} d\ell\tilde{\alpha}_{n-1}=\frac{1}{(n+1)!}\ell^{n+1}
\ea
\ee

From here we can find all other eventualities. The right-leaning
singlet clusters $e_n$ probabilities are:
\be
s_n=f_n=e_n=\int_0^{\ell}d\ell e^{-4\ell} \frac{l^{n+1}}{(n+1)!},
\ee
for $n\ge 0$, which is extremely simple considering the tortuous way to here. 
The final bin probability to end up in, say, bin $e_n$ is:
\be
e_n|_{\ell\rightarrow \infty}=\int_0^{\infty}d\ell e^{-4\ell}
\frac{l^{n+1}}{(n+1)!}=\frac{1}{4^{n+2}}.
\ee
The sum of all $e_n$ eventualities is:
\be
\summ_{n=0}^{\infty}e_n=\summ_{n=0}^{\infty}\frac{1}{4^{n+2}}=\frac{1}{12}
\label{edecay}
\ee
This is the probability to end up as a right-leaning singlet
cluster. The same calculation applies to the s-bins and f-bins. Indeed
- $6\times \frac{1}{12}=1/2$, which completes the probability
of $2e_{-1}$ to 1. Note that the total probability of the spin-1
surviving and becoming a new spin-1 is the total $s_n$ sum, which is
also $1/6$.

The $RG$ time of the process follows directly from the above
discussion. Let us calculate the total time it takes for a spin-1 to be
completely eliminated. This we do in a self consistent way: The time
it takes to get completely decimated from the $s_n$ outcomes, is the
same as the total RG time we are seeking. Thus:
\be
\overline{\ell}_1=2\int de_{-1}\cdot \ell +2\summ_{n=0}^{\infty}\int (df_n+de_n)\cdot \ell+2\summ_{n=0}^{\infty}\int ds_n\cdot (\ell+\overline{\ell}_1)
\ee
where the factor of 2 is due to the reflection symmetry of the
configuration enumeration.
Upon use of the definitions of $e_n,\,f_n,\,s_n$ we have:
\begin{widetext}
\be
\frac{5}{6}\overline{\ell}_1=2\int_0^{\infty} e^{-4\ell}\ell
d\ell+6\int_0^{\infty}e^{-4\ell}\summ_{n=0}^{\infty}
\frac{\ell^{n+2}}{(n+1)!}\vspace{2mm}\\
=\frac{1}{8}+6\int_0^{\infty}e^{-4\ell} \ell
\l(e^{\ell}-1\rr)=\frac{1}{8}+\frac{6}{9}-\frac{6}{16}=\frac{5}{12}
\ee
\end{widetext}
Hence:
\be
\overline{\ell}_1=\frac{1}{2}
\label{l1}
\ee
This is the RG time for a spin-1 partition-site to be decimated
completely, and form an uncorrelated partition-bond. 

\subsection{Getting from a (2,0) domain to a spin-1 site \label{spin20}}

A more complicated analysis is required for the calculation of the
history of a partition-bond between a Ma-Dasgupta decimation and the
formation of a spin-1 partition-site. We must consider
two possibilities: first, the bond being a (2,0) or (0,2) domain, or
second, being a (1,1) domain. The latter is simpler, hence we start
with the former. 

Right after a Ma-Dasgupta decimation, the bond distribution $R(\beta)$ is quite
different from $P(\beta)$. It is actually the convolution of
$P(\beta)$ with itself:
\be\ba{c}
Q(\beta)=P(\beta_1)\stackrel{\beta}\times
P(\beta_2)\vspace{2mm}\\
=\intt_0^{\infty}d\beta_1\intt_0^{\infty}d\beta_2
P(\beta_1) P(\beta_2)\delta_{(\beta_1+\beta_2-\beta)}=\frac{4}{\G^2}\beta
e^{-2\beta/\G}
\label{Qeq}
\ea
\ee
As the RG progresses, this distribution may rebound to be closer to
$P(\beta)$. The flow equation it obeys is:
\begin{widetext}
\be
\frac{dR(\beta)}{d\G}=\der{R(\beta)}{\beta}+2\cdot P(0)\frac{1}{2}P(\beta_1)\stackrel{\beta}\times
R(\beta_2)-P(0) R(\beta).
\label{Req}
\ee
\end{widetext}
The second term represents the probability that a bond on either of the two
sides (hence the factor 2) gets decimated (hence $P(0)$). If the two
bonds right next to the decimated bond are different (probability
half), then nothing happens to the bond on the partition - the decimated
bond either forms a new spin-1, or connects a (1,1) domain to a spin
on the side of the partition. But if the domains near the
decimated-side bonds are the same (probability $1/2$ as well), we get
a Ma-Dasgupta decimation that involves the partition bond, hence the
convolution. In this case, we also need to remove the original
probability from the distribution $R(\beta)$, which is the origin of
the last term. 

\begin{center}
\begin{figure*}
\includegraphics[width=15cm]{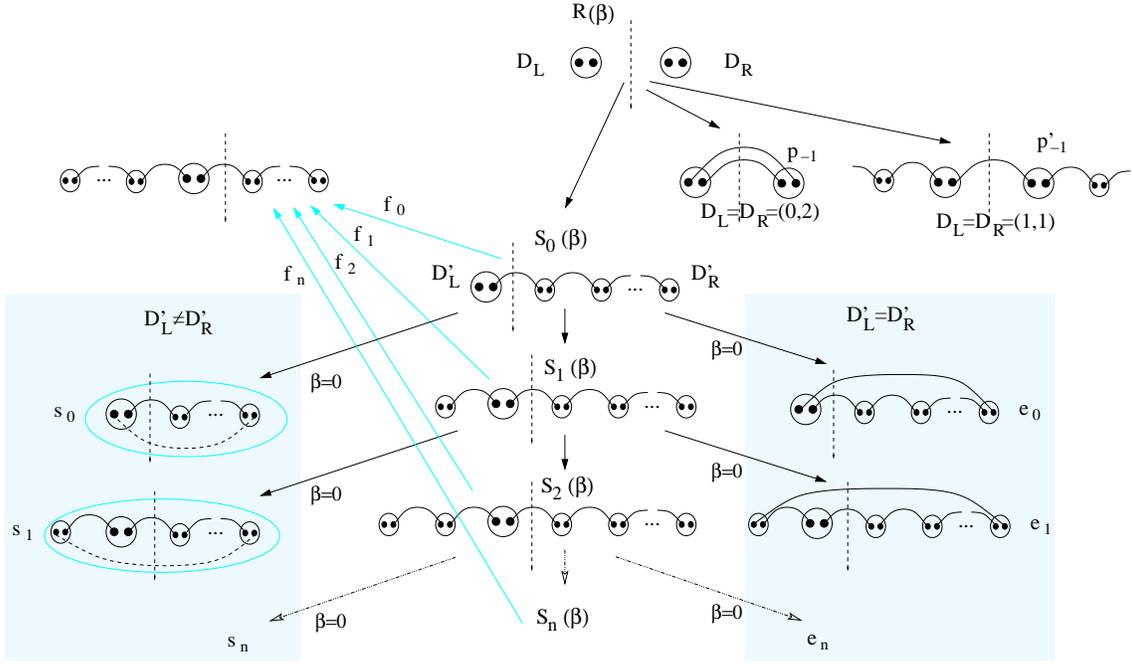}
\caption{The possible outcomes of a partition bond of domain
  (2,0). When the partition-bond gets decimated, and it is between two
  identical domains, $D_L=D_R$, then the Ma-Dasgupta rule is applied,
  Depending on whether $D_L=D_R=(1,1)$ or not, eventualities
  $p_{-1},\,p'_{-1}$ are obtained.
If $D_L\neq D_R$, when the partition-bond gets decimated it joins a (1,1) domain, in which case the
  bond strength has probability distribution $S_0(\beta)$. The (1,1) domain can be decimated by forming
  another singlet (eventuality $e_0$), by forming a triplet
  (eventuality $s_0$) or by joining on its left (right) another (1,1)
  domain through a Ma-Dasgupta decimation (eventuality $f_0$). It could also form another
  bond to its left, giving rise to the next line of diagrams, and the
  eventualities $S_1(\beta),\,e_1,\,s_1,\,f_1$. The subscript
  indicates the number of bonds to the left of the partition  that are
  involved in the (1,1) domain. Note that the $e_n$ and $s_n$
  eventualities occur when $D_L=D_R$ and $D_L\neq D_R$ respectively,
  and thus with the same probability due to domain-wall permutation
  symmetry ($D_{R/L}$ are the domains to right/left of the bond). 
\label{figRG2}}
\end{figure*}
\end{center}

The solution of Eq. (\ref{Req}) is straightforward, since $R(\beta)$ lives in the functional space spanned by $P(\beta)$ and
$Q(\beta)$. If we write:
\be
R(\beta)=a_{\ell}P(\beta)+b_{\ell}Q(\beta),
\label{Rform}
\ee
we obtain:
\be
\frac{d}{d\ell}\l(\ba{c} a_{\ell} \vspace{2mm}\\ b_{\ell} \ea \rr)=\l(\ba{cc} -3 &
2  \vspace{2mm}\\ 
1 & -2 \ea \rr)\l(\ba{c} a_{\ell} \vspace{2mm}\\ b_{\ell} \ea \rr)
\ee
The eigenvalues and eigenvectors of the matrix on the RHS are:
\be
\ba{cc}
-1 & (1,1)\vspace{2mm}\\
-4 &  (2,-1)
\ea
\ee
Given that the initial conditions are $a_{\ell=0}=0,\,b_{\ell=0}=1$,
we obtain:
\be
\ba{c}
a_{\ell}=\frac{2}{3}\l(e^{-\ell}-e^{-4\ell}\rr)\vspace{2mm}\\
b_{\ell}=\frac{1}{3}\l(2e^{-\ell}+e^{-4\ell}\rr)
\ea
\label{Rsol}
\ee

From here on, we can follow the same procedure as
Sec. \ref{spin1}. The first thing to consider is having the partition-bond be decimated, with the same domains on its sides. This invokes the
Ma-Dasgupta rule, terminates this step, and begins a
new cycle. The differential probability of this happening is $\frac{1}{2} R(0)d\G$. But there
are two sub-possibilities: the neighboring domains could be either
(1,1), or (2,0)/(0,2), each with probability $1/2$. The first feeds
into $p'_{-1}$, and the second to $p_{-1}$. So we obtain:
\be\ba{c}
\frac{dp_{-1}}{d\G}=\frac{dp'_{-1}}{d\G}\vspace{2mm}\\
=\frac{1}{4}R(0)=\frac{1}{4}\frac{2}{\G}\frac{2}{3}\l(e^{-\ell}-e^{-4\ell}\rr)
\ea
\ee
Thus:
\be
\frac{dp_{-1}}{d\ell}=\frac{dp'_{-1}}{d\ell}=\frac{1}{3}\l(e^{-\ell}-e^{-4\ell}\rr)
\ee
and:
\be
p_{-1}=p'_{-1}=\frac{1}{3}\l(1-e^{-\ell}-\frac{1}{4}(1-e^{-4\ell})\rr)
\ee
and as $\ell\rightarrow\infty$, we have:
\be
p_{-1}|_{\ell\rightarrow\infty}=p'_{-1}|_{\ell\rightarrow\infty}=\frac{1}{4}
\ee
So with probability $1/2$, our bond gets decimated via a Ma-Dasgupta
rule right away. 

The alternative possibilities arise from the case of the partition
bond being decimated while its neighboring domains are different from
each other - which happens with probability $1/2$. In this case the partition bond joins a (1,1) domain to
its right or to its left, each with another probability factor of
$1/2$. When this happens, we need to follow a distribution,
$S_0(\beta)$, which has the same meaning and flow equation as $S_0(\beta)$ from
Sec. \ref{spin1}, except for the source term, which is now
$\frac{1}{4}R(0)P(\beta)$ (the $P(\beta)$ arises since it is the
distribution of the (1,1) domain to which the partition-bond annexes):
\begin{widetext}
\be
\frac{dS_0(\beta)}{d\G}=\der{S_0(\beta)}{\beta}+\frac{1}{2}P(0)(S_0(\beta)+P(\beta_1)\stackrel{\beta}\times
S_0(\beta_2))-2P(0) S_0(\beta)+\frac{1}{4}R(0) P(\beta)
\label{S0eq-prime}
\ee
\end{widetext}

As in Sec. \ref{spin1}, once the partition enters the distribution
$S_0(\beta)$ three possibilities for termination of this stage exist:
(1) form a singlet configuration via the $e_n$ route, (2) be involved
in a Ma-Dasgupta decimation with the bond to the opposite side of the
(1,1) domain - $f_n$ route, and (3) the partition-bond can be
decimated while its neighboring domains are different, giving rise to
a spin-1 containing the partition - the $s_n$ route. Also,
$S_0(\beta)$ can flow into the additional $S_n(\beta)$ just as
before. 

It is easy to see that the flow equations for $e_n,\,f_n,\,s_n$ and
$S_n(\beta)$ are the same as Eqs. (\ref{neq}). The only difference is
$S_0(\beta)$, due to its different source term. Let us solve
$S_0(\beta)$ for this case. Here too, we can write
$S_0(\beta)=\alpha_0(\ell) P(\beta)$. Once we substitute this into
Eq. (\ref{S0eq-prime}), we obtain:
\be
\frac{d\alpha_0}{d\ell}=-4\alpha_0+\frac{1}{2}\frac{2}{3}(e^{-\ell}-e^{-4\ell})
\ee
As before, it is beneficial to follow
$\tilde\alpha_0=e^{4\ell}\alpha_0$. This obeys:
\be
\frac{d\tilde{\alpha}_0}{d\ell}=\frac{1}{3}(e^{3\ell}-1). 
\label{talpha0}
\ee

The next step is to find $\tilde{\alpha}_n$, which are the
probability 'amplitudes' of $S_n(\beta)$, as in
Eq. (\ref{alphaville}). They obey Eq. (\ref{talpha}). 
As it turns out, it is helpful to expand the exponent in
Eq. (\ref{talpha0}) into a power-law. Upon integration we obtain:
\be
\tilde{\alpha}_0=\frac{1}{3}\summ_{j=1}^{\infty}\frac{1}{3}\frac{(3\ell)^{j+1}}{(j+1)!}. 
\ee
from which we can immediately find:
\be
\tilde{\alpha}_n=\frac{1}{3}\summ_{j=1}^{\infty}\frac{1}{3^{n+1}}\frac{(3\ell)^{j+n+1}}{(j+n+1)!}
=\frac{1}{3^{n+2}}(e^{3\ell}- \summ_{j=0}^{n+1}\frac{(3\ell)^{j}}{j!})
\ee
From this result we can find $e_n=f_n=s_n$ as a function of $\ell$:
\be
\frac{de_n}{d\ell}=\tilde\alpha_n e^{-4\ell} 
\ee
and hence:
\be
e_n(\ell)=\intt_0^{\ell}d\ell
\frac{1}{3^{n+2}}\l(e^{-\ell}-e^{-4\ell}\summ_{j=0}^{n+1}\frac{(3\ell)^{j}}{j!}\rr)
\ee
If we carry the integration to $\ell\rightarrow\infty$ we obtain:
\be
\ba{c}
e_n(\infty)=\frac{1}{3^{n+2}}\l(1-\frac{1}{4}\summ_{j=0}^{n+1}\l(\frac{3}{4}\rr)^j\rr)\vspace{2mm}\\
=\frac{1}{3^{n+2}}\l(1-\frac{1}{4}\frac{1-(3/4)^{n+2}}{1-3/4}\rr)\vspace{2mm}\\
=\frac{1}{4^{n+2}}
\ea
\ee
The sum over all $e_n$ is just:
\be
\summ_{n=0}^{\infty}e_n=\frac{1}{16}\frac{1}{1-1/4}=\frac{1}{12}
\ee
just as before.

We are now ready to calculate RG times. The total RG time for this
stage, between a Ma-Dasgupta decimation and the formation of a
partition-site is:
\begin{widetext}
\be
\overline{\ell}_0=
2\int dp_{-1}\cdot \ell +6\summ_{n=0}^{\infty}\int
de_n\cdot
\ell=\frac{2}{3}(1-\frac{1}{16})+6\summ_{n=0}^{\infty}\intt_0^{\infty}d\ell\cdot\ell e^{-4\ell}\tilde{\alpha}_n
\ee
As it turns out, we can carry out the sum over $\tilde{\alpha}_n$
directly:
\be\ba{c}
\summ_{n=0}^{\infty}\tilde{\alpha}_n=\summ_{n=0}^{\infty}\l(\frac{e^{3\ell}}{3^{n+2}}-\summ_{j=0}^{n+1}\frac{(3\ell)^j}{j!}\rr)\vspace{2mm}\\
=\frac{1}{6}e^{3\ell}-\frac{1}{3}\summ_{j=0}^{\infty}\summ_{n=j-1}^{\infty}\frac{1}{3^{n+1}}\frac{(3\ell)^j}{j!}+1/3\vspace{2mm}\\
=\frac{1}{6}e^{3\ell}-\frac{1}{2}e^{\ell}+\frac{1}{3}.
\ea
\ee
And thus:
\be
\overline{\ell}_0=\frac{2}{3}(1-\frac{1}{16})+6\intt_0^{\infty}d\ell\ell \l(\frac{1}{6}e^{3\ell}-\frac{1}{2}e^{\ell}+\frac{1}{3}\rr)=\frac{5}{8}+1-\frac{1}{3}+\frac{1}{8}=\frac{17}{12}.
\label{l0}
\ee
\end{widetext}
This is the average time it takes to become either a Ma-Dasgupta
decimated bond, or a spin-1 containing the partition (probability $1/6$).

\subsection{Getting from a (1,1) domain to a spin-1 site}

\begin{center}
\begin{figure*}
\includegraphics[width=15cm]{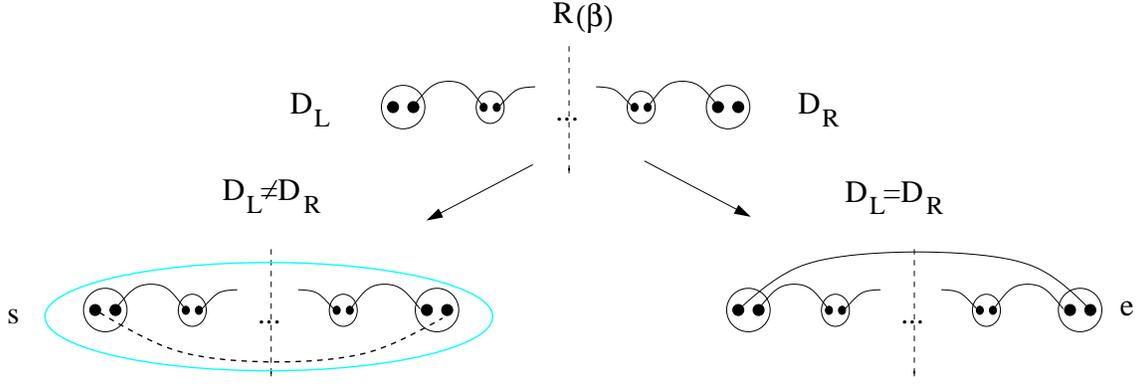}
\caption{The possible outcomes of a partition bond of domain
  (1,1). When the partition-bond gets decimated, if it is between two
  identical domains, $D_L=D_R$, then the Ma-Dasgupta rule is applied,
  otherwise, it forms a partition-spin-1.
\label{figRG3}}
\end{figure*}
\end{center}

A fresh partition-bond of a (1,1) domain
can terminate either through the formation of a full singlet, or by
the formation of a spin-1 partition-site (Fig. \ref{figRG3}).

As in Sec. \ref{spin20}, we first find the flow equation for the
distribution $R(\beta)$ of the partition-bond right after
decimation:
\begin{widetext}
\be
\frac{dR(\beta)}{d\G}=\der{R(\beta)}{\beta}+2\cdot P(0)\l(\frac{1}{2}P(\beta_1)\stackrel{\beta}\times
R(\beta_2)+\frac{1}{2}R(\beta)\rr)-2P(0) R(\beta).
\label{Req11}
\ee
\end{widetext}
Note that there are two differences with respect to the analog
Eq. (\ref{Req}) in Sec. \ref{spin20}. The brackets contain another term
$\frac{1}{2}R(\beta)$, which arises from the possibility of a bond to
either side of the partition-bond being decimated, but with a different
domain on its other side. This will extend the (1,1) domain of the
partition-bond, but will maintain its strength. The other difference
is the factor of 2 in the last term - this change is due to the fact
that now {\it any} decimation of a bond neighboring the partition bond
leads to its renormalization. In spite of these differences, the two
additions cancel, and we obtain exactly the same result for $R(\beta)$
as in Eqs. (\ref{Rform}) and (\ref{Rsol}). 

All that remains is to find the dependencies of the two probabilities
$e$ and $s$ on $\ell$, where $e$ is the probability of the
partition-bond being decimated with its domains being the same on both
sides, and $s$ is the probability of being decimated with different
domains surrounding the partition bond, and hence forming a spin-1
(see Fig. \ref{figRG3}). It easy to see that these probabilities are
equal, and that:
\be
\frac{de}{d\G}=\frac{1}{2}R(0)=\frac{1}{\G} \frac{2}{3}(e^{-\ell}-e^{-4\ell})
\ee
and hence:
\be
s(\ell)=e(\ell)=\frac{2}{3}(1-e^{-\ell}-\frac{1}{4}(1-e^{-4\ell}))
\ee
as $\ell\rightarrow\infty$ we indeed obtain $s(\infty)=e(\infty)=1/2$. 

The RG-time for this stage, i.e., between a Ma-Dasgupta decimation and
the decimation of the partition bond (either FM or AFM), is:
\be
\overline{\ell}_{11}=
2\int de \cdot \ell =2\intt_0^{\infty}d\ell \cdot \ell\cdot
\frac{2}{3}(e^{-\ell}-e^{-4\ell})=\frac{4}{3}\l(1-\frac{1}{16}\rr)=\frac{5}{4}.
\label{l11}
\ee
Note that now the probability of becoming a spin-1 (s eventuality) is 1/2.

\subsection{Total average time}

The {\it average total} RG time between two Ma-Dasgupta decimation is given
as the sum of all the time of the limited histories times their
probabilities. For instance, the total time it takes for a partition-bond
which is a (1,1) domain to be completely eliminated is:
\be
\ell_{(1,1)}=\overline{\ell}_{11}+\frac{1}{2}\overline{\ell}_1=\frac{5}{4}+\frac{1}{4}=\frac{3}{2}
\ee
The last term is the product of the probability of forming a
partition-site, and the time for this spin-1 to be completely
decimated. We use the results from Eqs. (\ref{l1}) and (\ref{l11}).

A similar result applies to the (2,0) or (0,2) initial domains:
\be
\ell_{(2,0)}=\ell_{(0,2)}=\overline{\ell}_{0}+\frac{1}{6}\overline{\ell}_1=\frac{17}{12}+\frac{1}{6}\cdot\frac{1}{2}=\frac{3}{2}
\ee
where we also use Eq. (\ref{l0}).
Since this result is the same as for the (1,1) domain, the {\it total
  average } RG-time is:
\be
\ell_{total}=\frac{3}{2}.
\ee

This number should be compared to the RG time between 
Ma-Dasgupta decimations in the random-singlet phase, which is:
\be
\ell_{total}^{RS}=3.
\ee
Note also that this calculation is a generalized return-to-the-origin
probability.   

%-----------------------------------------------------Reduced entropy
\section{Reduced entropy of the spin-1 chain \label{reduced}}

After analyzing the history dependence and probabilities of the
various singlet configurations, what remains is the analysis of the
amount of quantum entanglement in them. Because of the complexity of the
quantum entanglement of SBS configurations due to the correlations
they produce, it is helpful to define a measure of entanglement which
neglects these short range correlations. Such a measure would be a
direct count of the number of singlets that connect the two parts of
our chain, and can be calculated exactly. In this section we define
this {\it reduced entanglement entropy}, $E$, analyze its
properties, and calculate it both analytically, and in
Sec. \ref{numerics} numerically, thus verifying our analytical calculation.

\subsection{Definition of the reduced entropy}

The ground state of the random spin-1 Heisenberg model is constructed
through iterative decimations of strong AFM and FM bonds. The reduced
entanglement with respect to a partition is defined as follows: 
An SBS connecting between one side of the partition to the other
contributes reduced-entropy of 1. Some spin-1 sites are clusters that
are the result of the decimation of a FM bond. If the partition is
contained within such a cluster, a SBS that connects the partition
site with the sites to its right contributes a fraction $\sigma$ to
the reduced entropy, and an SBS to its left contributes a fraction
$1-\sigma$. 

$\sigma$ is supposed to capture the internal structure of the spin-1
partition-site, and in principle may vary from cluster to cluster. When two clusters combine, their weight $\sigma$ should be
the average of their weights,
$\sigma_{12}=\frac{1}{2}(\sigma_1+\sigma_2)$. But using reflection
symmetry of the problem, we now show that we can set 
\be
\sigma=1/2
\ee
without loss of generality. Every time that
an SBS forms between a partition-spin-1 to a site on the
right, we would have $E=\sigma$ per singlet. But because of the
reflection symmetry of the problem we also need to consider the reflected
configuration, which would have $E=(1-\sigma)$ per singlet. The average
of these contributions is $1/2$, independent of $\sigma$. 
Thus, for simplicity, we set $\sigma=1/2$. 

Since the reduced entropy $E$ neglects correlations between the
spin-1/2's degrees of freedom connected by the SBS's, it is a purely
additive quantity: the reduced entanglement of a configuration is
the sum of the reduced entanglement of the individual
singlets.

\subsection{Reduced entropy of a partition-site-1}

\begin{center}
\begin{figure*}
\includegraphics[width=15cm]{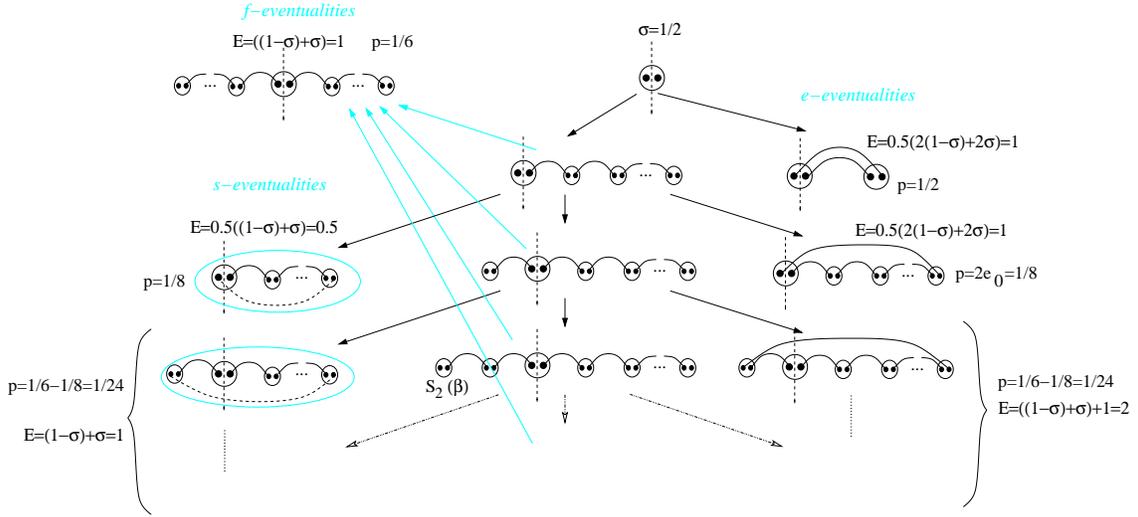}
\caption{Starting with a partition-spin-1, the various possibilities
  are illustrated on the same chart as Fig. \ref{figRG1}. $\sigma$ is
  the effective weight of the partition spin-1 that we begin
  with. Note that we omit here the future reduced entanglement in the
  case of the resulting partition-spin-1.
\label{figRG4}}
\end{figure*}
\end{center}

We start our analysis with a calculation of the average reduced
entanglement of configurations that result from a partition site. Once
we obtain our result for this intermediate stage we will consider the
average reduced entanglement of the beginning stages from a Ma-Dasgupta
decimation up to the formation of the spin-1 partition site. 

The entropy contribution and probabilities of a partition-spin-1 are
illustrated in Fig. \ref{figRG4}. The average contributions for this
stage, and its futures are (going from right to left in the figure):
\be
\ba{c}
E_{1}=\vspace{2mm}\\
\frac{1}{2}\cdot 1+ \frac{1}{8}\cdot 1+
\l(\frac{1}{6}-\frac{1}{8}\rr)\cdot 2+\frac{1}{6}\cdot
1+\frac{1}{8}\cdot\frac{1}{2}\vspace{2mm}\\
+\l(\frac{1}{6}-\frac{1}{8}\rr)\cdot 1+\frac{1}{6}E_1
\ea
\ee
where the last term self consistently adds the probability of forming
a partition-spin-1, which is $1/6$, and it having the same entropy
$E_1$. We get:
\be
E_1=\frac{47}{40}.
\ee

\subsection{Reduced entropy of a proton-bond of domain (2,0)}

\begin{center}
\begin{figure*}
\includegraphics[width=15cm]{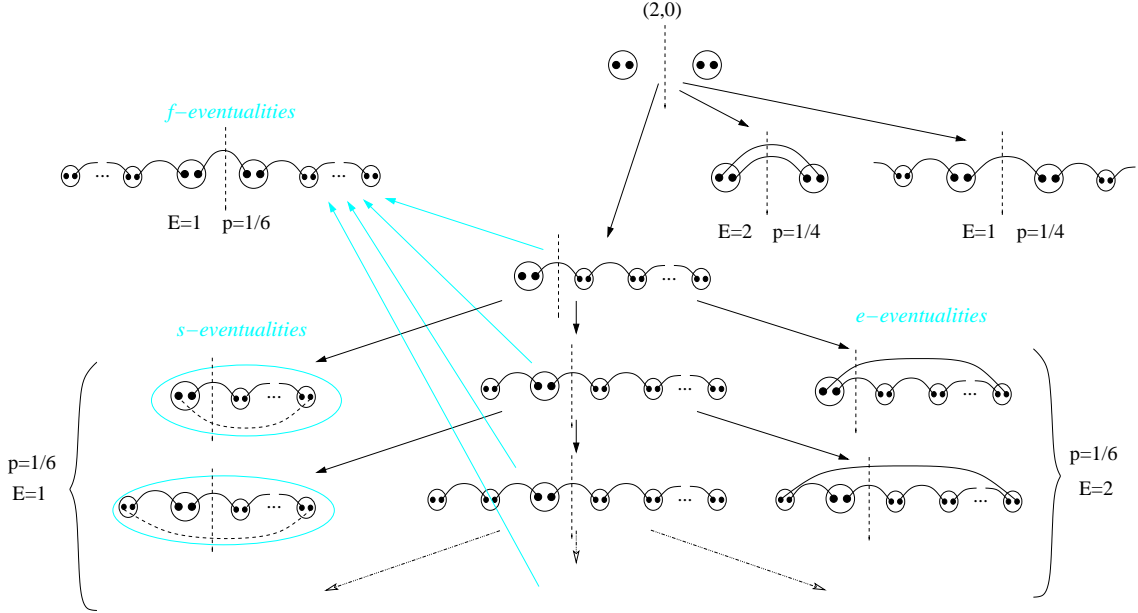}
\caption{Reduced entanglement starting with a (2,0) partition bond. The various possibilities
  are illustrated on the same chart as Fig. \ref{figRG2}. Note that we omit here the future reduced entanglement in the
  case of the resulting partition-spin-1.
\label{figRG5}}
\end{figure*}
\end{center}

The entropy contribution and probabilities of a partition-bond of a
(2,0) domain are illustrated in Fig. \ref{figRG5}. The average contributions for this
stage, together with the partition-spin-1 stage are: 
\be
\ba{c}
E_{(2,0)}=\vspace{2mm}\\
\frac{1}{4}\cdot 1+\frac{1}{4}\cdot 2+\frac{1}{6}\cdot
2+\frac{1}{6}\cdot 1+\frac{1}{6}\cdot
1+\frac{1}{6}E_1\vspace{2mm}\\
=\frac{17}{12}+\frac{47}{6\cdot 40}.
\ea
\ee

\subsection{Reduced entropy of a partition-bond of domain (1,1)}

\begin{center}
\begin{figure*}
\includegraphics[width=15cm]{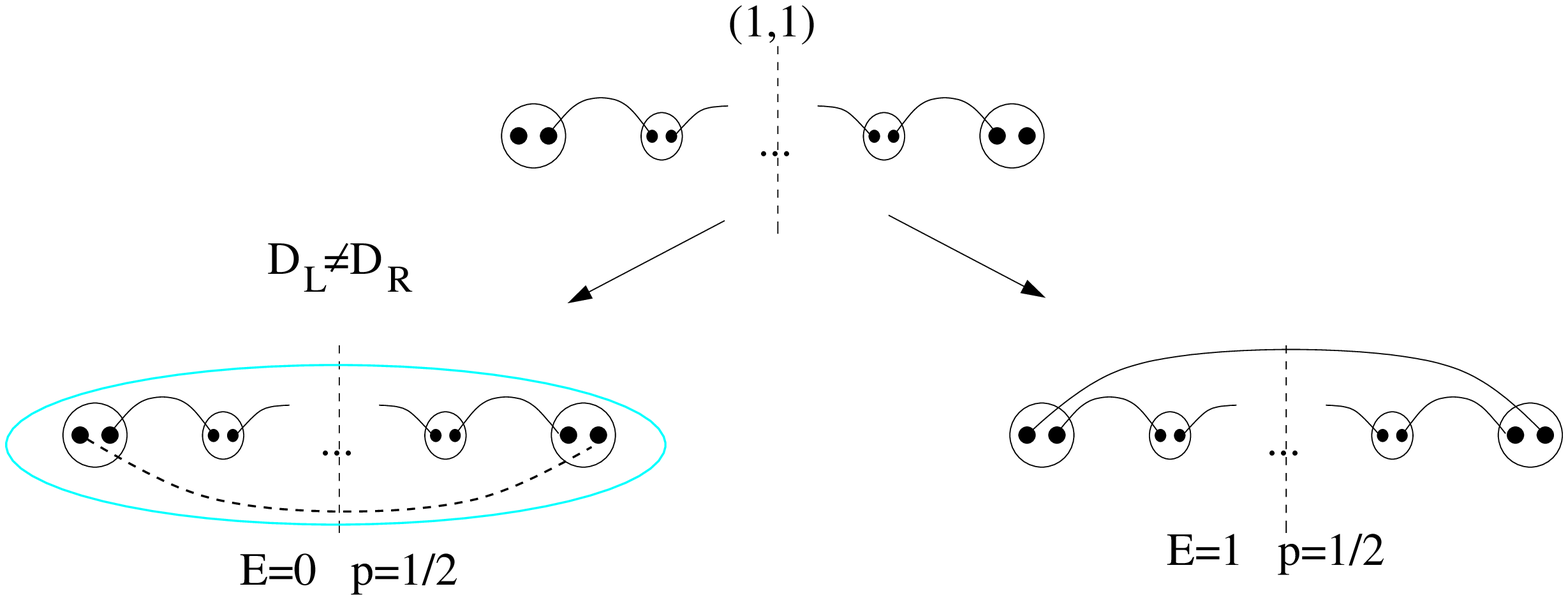}
\caption{Reduced entanglement starting with a (1,1) partition bond. The various possibilities
  are illustrated on the same chart as Fig. \ref{figRG3}. Note that we omit here the future reduced entanglement in the
  case of the resulting partition-spin-1.
\label{figRG6}}
\end{figure*}
\end{center}

The entropy contribution and probabilities of a partition-bond of domain(1,1) are
illustrated in Fig. \ref{figRG6}. Note that the original singlet of
the (1,1) domain does not contribute, since we considered it when it
formed. The average contributions for this
stage, together with the partition-spin-1 stage are: 
\be
E_{(1,1)}=\frac{1}{2}\cdot 1+\frac{1}{2}E_1=\frac{1}{2}+\frac{47}{2\cdot 40}
\ee

\subsection{Average total reduced entropy, and reduced central charge}

The average total reduced entropy is:
\be
\ba{c}
E_{total}=\frac{1}{3}E_{(1,1)}+\frac{2}{3}E_{(2,0)}\vspace{2mm}\\
=\frac{1}{6}+\frac{47}{6\cdot
  40}+\frac{17}{3\cdot 6}+\frac{47}{9\cdot
  40}\vspace{2mm}\\
=\frac{26}{18}-\frac{1}{8\cdot 18}
\ea
\ee

We are now in a position to calculate the reduced central charge of
the spin-1 critical point. The entanglement of a segment $L$ with the
rest of the chain consists of: 
\be
\ba{c}
E=2\cdot\frac{E_{total}}{\ell_{total}}\ell_L\vspace{2mm}\\
=2\frac{\frac{26}{18}-\frac{1}{8\cdot
18}}{3/2}\ln
L^{1/3}\vspace{2mm}\\
=\frac{1}{3}\frac{4}{3}\l(\frac{26}{18}-\frac{1}{8\cdot
18}\rr)\ln L\vspace{2mm}\\
=\frac{1}{3}\l(2-\frac{1}{12}\rr)\ln L 
\ea
\ee
Thus the
reduced entropy central charge is: 
\be c_r=\l(2-\frac{1}{12}\rr)\ln 2
\approx 1.917\ln 2 
\ee 
which should be compared to $c_r \approx 1.923
\ln 2$ found numerically in the following section.

\subsection{Numerical evaluation of the reduced entropy
  \label{numerics}}

The previous sections have described a method for tracking history
dependence in the RSRG in order to extract the universal part of the
entanglement entropy.  Since this method is fairly intricate, a direct
numerical check on the results is worthwhile.  Although the full
entanglement entropy cannot be calculated in closed form but only in a
controlled approximation, the ``reduced entropy'' above can be
calculated exactly.  As a check, we now compute the reduced entropy by
a numerical Monte Carlo simulation of the RSRG equations on a finite
spin-1 chain.

\begin{center}
\begin{figure}
\includegraphics[width=8cm]{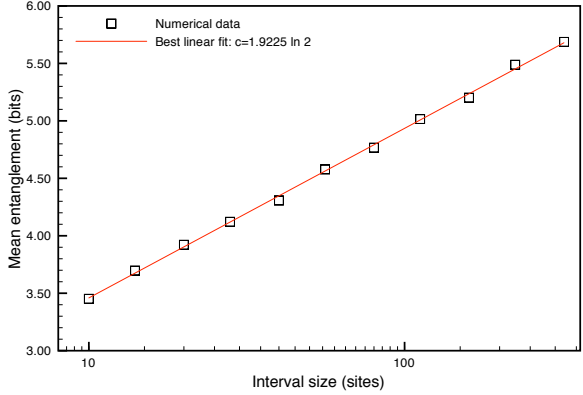}
\caption{Numerical calculation of the reduced entropy $S_{\rm red}$ in the $s=1$ chain for intervals of various sizes $N$ within a chain of length $10^4$ sites.  The fit is to $S_{\rm red} = S_0 + (c/3) \log_2 N$.  The statistical standard deviation for each point shown is less than $5\%$.  \label{numericalplot}} 
\end{figure}
\end{center}

For each subsystem size $N$, we average over 100 different intervals
within each of 100 realizations.  The initial distribution of
Heisenberg couplings is tuned to lie at the critical point of an
infinite chain.  Fig.~\ref{numericalplot} shows the result: the
reduced entropy of an interval of size $N$ is approximately \be
S(N)=1.9837+\left({1.923 \ln 2 \over 3}\right) \log_2 N.  \ee An
overall statistical error in the numerical central charge of
approximately $2 \%$ is expected.  The agreement of the numerical
value $1.923 \ln 2$ with the analytic result $(23/12) \ln 2$ can be
taken to verify the history dependence obtained above.  It appears
that systematic errors in the numerics resulting from finite interval
size and finite chain size ($10^4$ sites) are small.  The remaining
step is to use the same analysis of history dependence to obtain the
true entanglement.

%-------------------------------------configurational entropy  

\section{Calculation of configurational entanglement in the spin-1
  critical point \label{conf}}

The last step on our way to the entanglement entropy of the
spin-1 chain at its critical point is to understand the quantum entanglement
of each closed singlet configuration. Many such configurations arise in the decimation
process, but fortunately, for a high accuracy evaluation of the
entanglement entropy, only few classes of these configurations are
necessary. In this section we will discuss the entanglement in these
configuration classes, both analytically and numerically. 

\subsection{Simple configurations}

Let us first consider the configurations arising from a (2,0) domain,
Fig. \ref{figRG2}. The simplest eventualities involve a decimation of the
(2,0) domain while its two neighboring domains are identical. There
are two possibilities. In eventuality $p_{-1}$ the neighboring domains
are (0,2), and the decimation of the partition-bond involves forming
a complete singlet between the spin-1 sites to the left and right of the
partition:
\be
\ket{\psi_{\ell~r}}=\frac{1}{2}\l(\ad{\ell}\bd{r}-\bd{\ell}\ad{r}\rr)^2\ket{0_{\ell~r}}.
\ee
Since this is a Ma-Dasgupta decimation, the entanglement due to this
step is independent of future steps. The entanglement between the
left and the right sites is simply:
\be
S_{p_{-1}}=\log_2 3
\ee
since a trace of the left (or right) sites lead to an SO(3) symmetric
density matrix for the right (left) site, and the spin-1 is completely
undetermined, hence all three states contribute equally to the
entropy.  

The second possibility is that the (2,0) partition-bond is near
(1,1) domains, which is  a $p'_{-1}$ eventuality. In this case, the domain
walls surrounding the partition-bond are spin-1/2, and the decimation
gives rise to one SBS, which gives:
\be
S_{p'_{-1}}=1.
\ee
A spin-1/2 degree of freedom is determined by the SBS on either side
of the partition. 

If the partition bond gets decimated while it is surrounded by two
different domains, then the bond joins the (1,1) domain by the
formation of an SBS. Since the (1,1) domain is assumed to
be very large in the scaling limit, the domain walls on both its sides
are uncorrelated. Nevertheless, we know that the partition is at a
small finite distance to one of the (1,1) domain edges. In the next
stages of the RG, the (1,1) domain expands, until it gets decimated in
one of three ways. $f$ eventualities will result in the (1,1) domain
of the bond joining another (1,1) domain through a Ma-Dasgupta
decimation on its short side. This results in a closed cluster, with
the entanglement entropy:
\be
S_{f}=1.
\ee

A more difficult configuration occurs if the (1,1) domain containing
the partition bond is decimated between two identical neighboring
domains, while the partition bond is a distance $n$ from the edge of
the (1,1) domain - $e_n$ eventuality of Fig. \ref{figRG2}. In the absence
of any ferromagnetic decimations, the entanglement of such a cluster
can be calculated exactly, using the $SO(3)$ symmetry, and an
important simplification. This simplification serves as a key
ingredient of the numerical calculation of the entanglement entropy of
more complicated configurations, that do involve ferromagnetic
decimations. This is described in the next section. 

\begin{center}
\begin{figure}
\includegraphics[width=8cm]{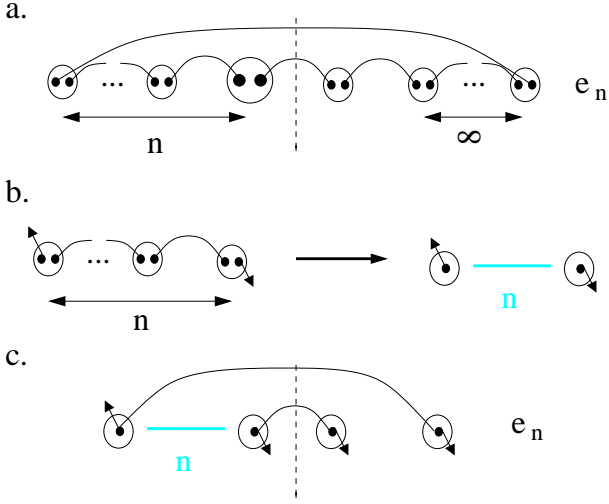}
\caption{(a) A typical e-eventuality configuration. (b) Using the
  chain-rule singlet row reduction, we can reduce a chain of $n+1$
  spin-1 sites with two dangling spin-1/2's, to just two exact
  spin-1/2's, which are related to the dangling edge spin-1/2's as in
  Eq. (\ref{rowtrans}). (c) Application of the chain-rule to the
  configuration in (a). \label{chainrule}} 
\end{figure}
\end{center}

\subsection{Singlet-row reduction \label{srr}}

To simplify the understanding of configurations such as in
Fig. \ref{chainrule}a, we will derive a rule that will replace the
Schwinger-boson singlet chains which are completely contained in one
side of the partition. Consider a chain of $n+1$ spin-1 sites, whose
state consists of a singlet-row, i.e., a single connects sites $m$ and
$m+1$ for all $m=0,\ldots,n-1$. Sites $0$ and $n$ have a dangling
spin-1/2, thus there are four singlet-row states, which we write as:
\be
\ba{c}
\ket{\uparrow~(1,1)~\uparrow}=
\ad{0}\ad{n}\prodd_{i=0}^{n-1}\l(\ad{i}\bd{i+1}-\bd{i}\ad{i+1}\rr)\ket{0}\vspace{2mm}\\
\ket{\uparrow~(1,1)~\downarrow}=
\ad{0}\bd{n}\prodd_{i=0}^{n-1}\l(\ad{i}\bd{i+1}-\bd{i}\ad{i+1}\rr)\ket{0}\vspace{2mm}\\
\ket{\downarrow~(1,1)~\uparrow}=
\bd{0}\ad{n}\prodd_{i=0}^{n-1}\l(\ad{i}\bd{i+1}-\bd{i}\ad{i+1}\rr)\ket{0}\vspace{2mm}\\
\ket{\downarrow~(1,1)~\downarrow}=
\bd{0}\bd{n}\prodd_{i=0}^{n-1}\l(\ad{i}\bd{i+1}-\bd{i}\ad{i+1}\rr)\ket{0}.
\ea
\label{4s}
\ee
The dangling spin-1/2's at the edge are eventually locked into
singlets, or ferromagnetic bonds which stretch past the partition. The sites $i=1,\,\ldots,\,n-1$ in
the singlet row are inert, and do not contribute to
entanglement, therefore calculation of the entanglement with them
included is going to be quite difficult since they increase the Hilbert
space involved by a factor of $3^{n-1}$. They do have an important
role, however, which is to facilitate the correlations along the
singlet chain. 

But the correlations of the singlet-row can be taken into account while
removing the interior inert sites. 
The correlations carried in the states in Eq. (\ref{4s}) are exhibited by the fact that
the four states are not orthogonal. In particular:
\be
\langle
\uparrow~(1,1)~\downarrow\ket{\downarrow~(1,1)~\uparrow}=(-1)^n
\label{overlap}
\ee
This overlap is easily found, since it is the product of the amplitude
of the state $m^z_i=0$ for all sites $0\le i\le n$. Note that the
states in Eq. (\ref{4s}) are not normalized:
\be
\ba{c}
\langle
\uparrow~(1,1)~\uparrow\ket{\uparrow~(1,1)~\uparrow}=\langle
\downarrow~(1,1)~\downarrow\ket{\downarrow~(1,1)~\downarrow}=N_{\uparrow\uparrow}\vspace{2mm}\\
\langle
\uparrow~(1,1)~\downarrow\ket{\uparrow~(1,1)~\downarrow}=\langle
\downarrow~(1,1)~\uparrow\ket{\downarrow~(1,1)~\uparrow}=N_{\uparrow\downarrow}
\ea
\label{rowtrans}
\ee
Let us now define four
new states that will be mutually orthogonal:
$\ket{\uparrow\uparrow},\,\ket{\uparrow\downarrow},\ket{\downarrow\uparrow},\,\ket{\downarrow\downarrow}$.
We can capture the correlations in the singlet-row by writing:
\be
\ba{c}
\ket{\uparrow~(1,1)~\uparrow}=\sqrt{N_{\uparrow\uparrow}}\ket{\uparrow\uparrow}\vspace{2mm}\\
\ket{\downarrow~(1,1)~\downarrow}=\sqrt{N_{\uparrow\uparrow}}\ket{\downarrow\downarrow}\vspace{2mm}\\
\ket{\downarrow~(1,1)~\uparrow}=a\ket{\downarrow\uparrow}+b\ket{\uparrow\downarrow}\vspace{2mm}\\
\ket{\uparrow~(1,1)~\downarrow}=a\ket{\uparrow\downarrow}+b\ket{\downarrow\uparrow}
\label{new4s}
\ea
\ee
where $a$ and $b$ are real numbers which obey:
\be
\ba{c}
a^2+b^2=N_{\uparrow\downarrow},\vspace{2mm}\\
2ab=(-1)^n,
\ea
\label{abeq}
\ee
which is solved by:
\be
\ba{c}
a=\frac{1}{2}\l(\sqrt{N_{\uparrow\downarrow}+(-1)^n}+\sqrt{N_{\uparrow\downarrow}-(-1)^n}\rr),\vspace{2mm}\\
b=\frac{1}{2}\l(\sqrt{N_{\uparrow\downarrow}+(-1)^n}-\sqrt{N_{\uparrow\downarrow}-(-1)^n}\rr).
\ea
\label{abdef}
\ee
The norms $N_{\downarrow\downarrow},\,N_{\uparrow\downarrow}$ are found
  in App. \ref{appA} to be:
\be
\ba{c}
N_{\uparrow\uparrow}=\frac{1}{2}\l(3^{n+1}+(-1)^n\rr),\vspace{2mm}\\
N_{\uparrow\downarrow}=\frac{1}{2}\l(3^{n+1}-(-1)^n\rr).
\ea
\ee

The eigenvalues of the reduced density matrix of states that consist of singlet-rows written just using the dangling spins, as
in Eq. (\ref{new4s}), will be the same as those for the reduced
density matrix involving all inert spin-1 sites, since the inner
product of the states in Eq. (\ref{4s}) and (\ref{new4s}) are the
same. This presents an extreme simplification in terms of numerical
evaluation of these eigenvalues. 

\subsection{Entanglement of the level-0 (no FM clusters) e-eventualities and f-eventualities\label{l0sec}}

Using the reduction of singlet-rows as in Eq. (\ref{new4s}) we can
reduce the configuration in Fig. \ref{chainrule}a to that of
Fig. \ref{chainrule}c. The entanglement entropy of the $e_n$
eventuality is then easily calculated. 

The first step is to write the state in Fig. \ref{chainrule}a while
ignoring the correlations along the singlet-row:
\be
\ba{c}
\ket{\psi}=\l(\ket{\uparrow_1}\ket{\downarrow_4}-\ket{\downarrow_1}\ket{\uparrow_4}\rr)\l(\ket{\uparrow_2}\ket{\downarrow_3}-\ket{\downarrow_2}\ket{\uparrow_3}\rr)\vspace{2mm}\\
=\ket{\uparrow_1}\ket{\downarrow_4}\ket{\uparrow_2}\ket{\downarrow_3}-\ket{\uparrow_1}\ket{\downarrow_4}\ket{\downarrow_2}\ket{\uparrow_3}\vspace{2mm}\\
-\ket{\downarrow_1}\ket{\uparrow_4}\ket{\uparrow_2}\ket{\downarrow_3}+\ket{\downarrow_1}\ket{\uparrow_4}\ket{\downarrow_2}\ket{\uparrow_3}.
\ea
\ee
Sites 1 and 2 are actually representatives of the dangling spin-1/2's
in the singlet-row, and therefore we apply to them the assignment of
Eq. (\ref{new4s}):
\be
\ba{c}
\ket{\psi}\rightarrow
\sqrt{N_{\uparrow\uparrow}}\ket{\uparrow_1}\ket{\downarrow_4}\ket{\uparrow_2}\ket{\downarrow_3}\vspace{2mm}\\
-a\ket{\uparrow_1}\ket{\downarrow_4}\ket{\downarrow_2}\ket{\uparrow_3}-b\ket{\downarrow_1}\ket{\uparrow_4}\ket{\downarrow_2}\ket{\uparrow_3}\vspace{2mm}\\
-a\ket{\downarrow_1}\ket{\uparrow_4}\ket{\uparrow_2}\ket{\downarrow_3}-b\ket{\uparrow_1}\ket{\downarrow_4}\ket{\uparrow_2}\ket{\downarrow_3}\vspace{2mm}\\
+\sqrt{N_{\uparrow\uparrow}}\ket{\downarrow_1}\ket{\uparrow_4}\ket{\downarrow_2}\ket{\uparrow_3}
\label{rotpsi}
\ea
\ee
with $a$ and $b$ defined as in Eq. (\ref{abdef}). The entanglement
entropy follows from the eigenvalues of the reduced density matrix of
the state $\ket{\psi}$ over sites 3 and 4. From SO(3) symmetry we can
argue that the form of the reduced density matrix is:
\be\ba{c}
\hat{\rho}_{12}={\rm
  tr_{34}}\hat{\rho}=\vspace{2mm}\\
\frac{1}{3}\alpha\l(\ket{\uparrow\uparrow}\bra{\uparrow\uparrow}+\ket{\downarrow\downarrow}\bra{\downarrow\downarrow}+ \frac{1}{2}\ket{\uparrow\downarrow+\downarrow\uparrow}\bra{\uparrow\downarrow+\downarrow\uparrow}\rr)\vspace{2mm}\\
+\beta\frac{1}{2}\ket{\uparrow\downarrow-\downarrow\uparrow}\bra{\uparrow\downarrow-\downarrow\uparrow}
\ea
\ee
where this representation breaks the reduced density matrix into the
triplet and singlet sectors respectively, where it is already
diagonal. We also have $\alpha=1-\beta$. To find $\alpha$ and $\beta$
all we need is to know the matrix element of one of the triplet states
in $\hat{\rho}_{12}$. This is easily found:
\be
\frac{1}{3}\alpha=\frac{N_{\uparrow\uparrow}}{2N_{\uparrow\uparrow}+2a^2+2b^2}=\frac{1}{2}\frac{N_{\uparrow\uparrow}}{N_{\uparrow\uparrow}+N_{\uparrow\downarrow}}
\ee
where the denominator is due to the normalization of the wave function
in Eq. (\ref{rotpsi}), and we used Eq. (\ref{abeq}). This is then:
\be
\alpha=\frac{3}{4}+\frac{(-1)^n}{4\cdot 3^n}
\label{alpha0}
\ee
The entanglement entropy of this configuration is then:
\begin{widetext}
\be
S_{e_n}^{(0)}=-\alpha\log_2\frac{\alpha}{3}-(1-\alpha)\log_2(1-\alpha)=-\l(\frac{3}{4}+\frac{(-1)^n}{4\cdot
  3^n}\rr)\log_2\l(\frac{1}{4}+\frac{(-1)^n}{4\cdot
  3^{n+1}}\rr)-\l(\frac{1}{4}-\frac{(-1)^n}{4\cdot
  3^n}\rr)\log_2\l(\frac{1}{4}-\frac{(-1)^n}{4\cdot 3^{n}}\rr)
\label{Sen}
\ee
\end{widetext}

The entanglement for the f-eventualities is remarkably simple, since
in this case there is a spin-1/2 singlet crossing the partition, but
this singlet is uncorrelated with the domain walls on the left and
right of the partition bond. The configurational entanglement of an
f-eventuality is:
\be
S_{f_n}^{(0)}=1.
\ee

\begin{center}
\begin{figure}
\includegraphics[width=8cm]{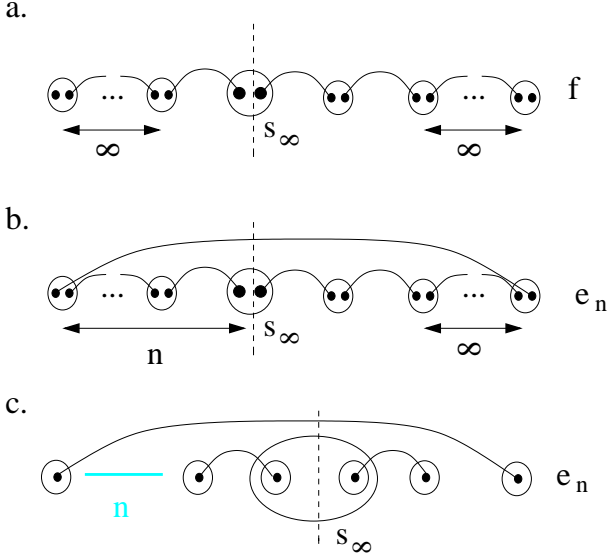}
\caption{(a) A typical f-eventuality following a FM decimation. The
  entanglement entropy across the partition-site can be calculated
  exactly if we neglect correlations between the two spin-1/2
  components of the partition site, a state which is denoted by $s_{\infty}$
(b) A typical e-eventuality configuration after a FM
  decimation. (c) Using the row-reduction rule we reduce the
  configuration from a to one of only six spin-1/2's. Neglecting
  internal correlations in the partition site, we can calculate this
  entanglement entropy exactly. \label{sinfty1}} 
\end{figure}
\end{center}

\subsection{Entropy of level-1 eventualities neglecting history
  dependence of cluster \label{l1sec}}

An approximate analytical answer for the entanglement can be obtained
using the self-consistency method already at level-1, i.e., considering configurations with at most one FM decimations, by
assuming that whenever a spin-1 cluster forms, its structure corresponds
to $s_{\infty}$ of Figs. \ref{figRG1} and \ref{figRG2}. This implies that the two spin-1/2's
participating in the ferromagnetic cluster are uncorrelated with the
partition bond, and are infinitely removed from it. In this case the
entropy contribution from level-n and level-n+1 become independent,
and the entropy can be summed self-consistently, as described in
Sec. \ref{self-consist}. 

For this purpose we need to derive the configurational entanglement of
$e_n$ and f eventualities of spin-1 sites, as in Fig. \ref{figRG1},
assuming the initial FM spin-1 cluster consists of two uncorrelated
spin-1/2 sites. Let us
begin with the simpler f-eventuality. This case is depicted in
Fig. \ref{sinfty1}a. The entanglement of this configuration, with
the spin-1 partition-site containing two spin-1/2's which are
uncorrelated before they form the FM cluster, is
easily computed to be:
\be
S_f^{(1)}=2-\frac{1}{2}\log_2 3.
\label{sfinfty}
\ee
Next we find the entanglement of the $e_n$ eventualities of this
setup, depicted in Fig. \ref{sinfty1}b. This setup is simplified to a
six-site configuration, with the partition situated in the
middle (Fig. \ref{sinfty1}c). Once the three spins to the right of the partition are traced
out, we are left with a 3-spin-1/2 density matrix, which for brevity
we do not quote here, of a complicated
mixed state. Due to rotational SO(3) symmetry, however, this density
matrix breaks down to invariant sectors corresponding to the
decomposition:
\be
1/2\times 1/2 \times 1/2=3/2+1/2+1/2.
\ee
One sector is the spin-3/2 subspace, and two other invariant subspaces will correspond to the
total spin of the three sites being 1/2. Furthermore the restriction
of the reduced density matrix on each of the invariant subspaces,
should be of diagonal form. In the basis $\ket{3/2,m}
(m=3/2,\ldots,-3/2),\,\ket{1/2,m} (m=1/2,-1/2), \,\ket{1/2',m}
(m=1/2,-1/2)$ the reduced density matrix is:
\be
\hat{\rho}_{1,2,3}={\rm diag}\{\alpha_n,\,\alpha_n,\,\alpha_n,\,\alpha_n,\,\beta_n,\,\beta_n,\,\gamma_n,\,\gamma_n\}
\ee
Finding $\alpha,\,\beta$, and $\gamma$ can be done
analytically. $\alpha$ is the easiest to obtain, since we know from
rotational symmetry that the state $\ket{\uparrow_1\uparrow_2\uparrow_3}$ is an
eigenstate.  We readily obtain:
\be
\alpha_n=\frac{1}{24}\l(1-(-1/3)^n\rr)
\label{esna}
\ee
Obtaining $\beta_n$ is more involved, but can be done by looking at the
restriction of the reduced density matrix to the invariant subspace of
$m_z=1/2$
($\ket{\downarrow_1\uparrow_2\uparrow_2},\,\ket{\uparrow_1\downarrow_2\uparrow_3},\,\ket{\uparrow_1\uparrow_2\downarrow_3}$),
and removing from it the subspace of
$\ket{3/2,m_z=1/2}=\frac{1}{\sqrt{3}}\l(\ket{\downarrow_1\uparrow_2\uparrow_2}+\ket{\uparrow_1\downarrow_2\uparrow_3}+\ket{\uparrow_1\uparrow_2\downarrow_3}\rr)$.
  This leaves us with an easily diagnosable $2\times 2$ matrix with
  eigenvalues:
\begin{widetext}
\be
\ba{c}
\beta_n=\frac{1}{24}\l(5+(-1/3)^n+\sqrt{\l(4+2\cdot (-1/3)^n\rr)^2-24
  \cdot(-1/3)^n+24/3^{2n}}\rr)\vspace{2mm}\\
\gamma_n=\frac{1}{24}\l(5+(-1/3)^n-\sqrt{\l(4+2\cdot (-1/3)^n\rr)^2-24
  \cdot(-1/3)^n+24/3^{2n}}\rr)
\ea
\label{esn}
\ee
\end{widetext}

The above formulas, Eqs. (\ref{esna}) and (\ref{esn}) obtain for
$n>1$. The cases of $n=-1$, $n=0$, and $n=1$ need to be calculated separately,
but their contribution to the entanglement entropy is very
simple. The cases of $e_{-1}$ and $e_0$ implies the spin-1 partition site is decimated
via two singlets that link it to two uncorrelated spin-1/2's to one of
its sides. Upon tracing out of that side of the partition, we obtain a
reduced matrix that describes a spin-1/2 with equal probability of
pointing up or down, hence, $e_0^{(1)}=e_{-1}^{(1)}=1$.

An analysis of the $e_1$
case can proceed along similar lines to the one above and to the
analysis of the $e_n$ eventualities as in Eq. (\ref{alpha0}), to
obtain:
\be
e_0^{(1)}=4\cdot \frac{1}{18}\log_2 18+2\cdot\frac{7}{18}\log_2
\frac{18}{7}
\ee

\begin{widetext}
Thus the entanglement of an $e_n$ cluster containing a spin-1
partition-site is:
\be
S_{e_n}^{(1)}=\l\{\ba{cr}
1 & n=-1,1\vspace{2mm}\\
4\cdot \frac{1}{18}\log_2 18+2\cdot\frac{7}{18}\log_2
\frac{18}{7} & n=1\vspace{2mm}\\
-4\alpha_n\log_2
\alpha_n-2\beta_n\log_2 \beta_n-2\gamma_n\log
_2 \gamma_n  & n>1.
\ea\rr.
\label{seinfty1}
\ee
\end{widetext}

\section{Entanglement entropy of the spin-1 chain \label{resultsec}}

We are now at a position to combine the above results into the
entanglement entropy, as given in Eqs. (\ref{Sintro}) and
(\ref{STintro}), to obtain:
\be
c_{eff}=\frac{2\overline{S}_{total}}{\overline{\ell}}\ln 2. 
\ee

In Sec. \ref{history} we found that:
\be
\overline{\ell}=\frac{3}{2}.
\ee
In the remains of this section, we will extract the value of
\be
\S_{total}=\summ_{c} p_c S_c,
\ee
where $p_c$ is the probability for a configuration $c$ to occur, and
$S_c$ is the entanglement entropy associated with it. 

One major obstacle is presented by the ferromagnetic decimations,
which give rise to complex quantum states. A decimated configuration can
involve any number, $n$, of spin-1 ferromagnetic decimation steps, but
with a rapidly decaying probability of $1/6^n$ (see Eq. \ref{edecay}). The
configurational entropy $\overline{S}_{total}$ can be obtained to
arbitrary accuracy by calculating exactly the configurational-entropy
of all configurations with $n-1$ ferromagnetic decimations, and
applying the approximation that the $n$-th level Ferromagnetic
decimation is between two spin-1/2's that are infinitely far from the
partition, and thus correspond to the $s_{\infty}$ eventuality. This
provides a result for $\overline{S}_{total}$ which has accuracy of:
\be
\Delta S_{total}\approx \frac{1}{6\cdot 6^n}.
\ee
In what follows we obtain the computer-free $n=1$ result, and using
Mathematica to calculate and sum the configurational entropy, we go up to
level $n=3$, with accuracy of $\sim 10^{-3}$.  

\subsection{Entanglement entropy from level 0 and 1 \label{self-consist}}

The entropy of the configurations computed exactly in Sec. \ref{l0sec}
and \ref{l1sec} can be summed up to give the $n=1$ approximation for
$\overline{S}_{total}$. In this approximation we have:
\be
\S_{total}=\frac{2}{3}\S_{(2,0)}+\frac{1}{3}\S_{(1,1)},
\label{ST1}
\ee
where $\S_{(2,0)}$ and $\S_{(1,1)}$ are the
average entanglement entropies across partition-bonds that are
initially (2,0) and (1,1) domains respectively. 

Let us first find the average entanglement with the partition bond
being a (2,0) domain initially. Looking at Fig. \ref{figRG2} we have:
\be
\S_{(2,0)}=\S_{(2,0)~e}^{(0)}+\S_{(2,0)~f}^{(0)}+\S_{(2,0)~s}^{(0)},
\ee
where $S_c^{(0)}$ is the average entanglement entropy due to
configuration $c$ with $c=e,\,f,\,s$. 
The entanglement due to the level-0 e-eventualities starting with a (2,0) or
(0,2) domain is given by:
\be
\S_{(2,0)~e}^{(0)}= p_{p'_{-1}}\cdot 1+p_{p'_{-1}}\cdot\log_2
    3+\summ_{n=0}^{\infty}S_{e_n}^{(0)}\cdot \frac{1}{8\cdot 4^n},
\label{Sezero}
\ee
where $p_{p'_{-1}}=p_{p_{-1}}=1/4$.

The entanglement due to level-0 f-entanglement starting with a (2,0)
domain is:
\be
\S_{(2,0)~f}^{(0)}=\summ_{n=0}^{\infty}S_{f_n}^{(0)}\cdot \frac{1}{8\cdot 4^n}=\frac{1}{6}.
\ee
To complete the calculation we need to add to these the entanglement
of the spin-1 cluster that forms due to the s-eventualities,
$\S_{(2,0)~s}^{(0)}$. In this section we approximate the spin-1 cluster as
consisting of uncorrelated spins. This produces average configurational
entropy we denote as $\S_{s}^{\infty}$, and thus:
\be
\S_{(2,0)~s}^{(0)}\approx p_s(1+\S_{s}^{\infty})=\frac{1}{6}(1+\S_{s}^{\infty}),
\ee
where the one in the brackets is due to the SBS formed on the
partition-bond in the process.

The equivalent contributions to level-0 when the partition-bond is a
(1,1) domain initially is quite simple. With probability $1/2$ the
partition bond undergoes a Ma-Dasgupta decimation which has
configurational entropy 1, which implies:
\be
\S_{(1,1)~e}^{(0)}=1/2\cdot 1=1/2.
\ee
The remaining probability is that the partition-bond is decimated
ferromagnetically. Thus:
\be
\S_{(1,1)}=\S_{(1,1)~e}+\S_{(1,1)~s}^{(0)}\approx \frac{1}{2}+\frac{1}{2}\S_s^{\infty}.
\ee

We now calculate $\S_s^{\infty}$. Looking at Fig. \ref{figRG1}, we see that:
\be
\S_{s}^{\infty}= \S_{e}^{(1)}+\S_f^{(1)}+\S_{s}^{(1)}.
\ee
The entanglement of an $e_n$ cluster containing a spin-1
partition-site is:
\begin{widetext}
\be
\S_{e~\infty}^{(1)}=\frac{5}{8}\cdot 1+\frac{1}{32}
e_0^{(1)}+\summ_{n=2}^{\infty}\frac{1}{8\cdot 4^n}\l(-4\alpha_n\log_2
\alpha_n-2\beta_n\log_2 \beta_n-2\gamma_n\log
_2 \gamma_n\rr).
\label{seinfty}
\ee
\end{widetext}
where we make use of the configurational entanglement we find in
Sec. \ref{l1sec}, and where the $\alpha_n,\,\beta_n,\,\gamma_n$ are
defined in Eqs. (\ref{esna}) and (\ref{esn}). 

$\S_{f}^{(1)}$ is much simpler, and from Eq. (\ref{sfinfty}), which
gives the configurational entropy, we get:
\be
\S_{f~\infty}^{(1)}=\frac{1}{6}(2-\log_2 3).
\ee

After considering all e- and f- eventualities, only one more
possibility remains, which is the formation of a new spin-1 cluster
via a ferromagnetic decimation. If we once more pursue the
approximation of assuming that the spin-1 partition-site forms out of
two spin-1/2's that are infinitely far from the partition, then we can
self consistently connect the entanglement of this eventuality with
the one calculated in this section. Thus we write:
\be
\S_{s}^{\infty}=\S_{f~\infty}^{(1)}+\S_{e~\infty}^{(1)}+\frac{1}{6}\S_{s}^{\infty}.
\ee
Isolating $\S_s^{\infty}$, and plugging in all numbers from Eqs. (\ref{sfinfty}) - (\ref{seinfty}) we
obtain the approximate cluster entropy due to a spin-1 partition
site:
\be
\S_{s}^{\infty}= 1.33495.
\label{Ssres}
\ee
Note that this is an {\it overestimate} since we neglect correlations
within the spin-1 effective site. These correlations reduce the
entanglement entropy. 

Combining the results above, and substituting into Eq. (\ref{ST1}) we
obtain that the average configurational entanglement is:
\be
\S_{total}=\frac{1}{3}\cdot 
1.16748+\frac{2}{3}\cdot 1.48271=1.3776.
\label{level0}
\ee

\subsection{Entanglement entropy from level 3 self consistent approximation}

In principle, we could obtain an exact result for the configurational
entanglement by summing up the entanglement of each configuration
times its probabilities. The probabilities of each and every
configuration are well known, but at levels higher than one FM
decimation, we can only obtain the configurational entropy using a
computer. Nevertheless, this part of the calculation can be
automated. 

The exact solution can be expressed using the following pattern. We can encode
the eventualities of decimated configurations as:
\be
\pi_0\circ\pi_1\circ\pi_2\circ\ldots\circ \pi_p,
\ee
where the $\pi_i$ describe what happens to the cluster at each level
of FM decimation. A series can terminate at
$\pi_0=p_{-1},\,p'_{-1},\,e_n^{R/L},\,f_n^{R/L}$, or begin by forming
a FM spin-1 partition site with $\pi_0=s_n^{R/L}$. The following p-1
steps can also be any sequence of FM spin-1 site formation,
$\pi_i=s_{n_i}^{R/L}$. The last step, $\pi_p$, can be any of
$\pi_p=e_{n_p}^{R/L},\,f_{n_p}^{R/L}$, which renders the cluster
decimated. Note that at levels $p>0$, $e_{n_p}$ has
$n_p=-1,\,0,\,1,\ldots$. 

Hence, an exact calculation can be written as:
\be
\S_{total}=\summ_{p=0}^{\infty}
\summ_{\{\pi_i|i=0,\,\ldots,\,p\}}\l(\prodd_{i=0}^{p} P_{\pi_i}\rr)
S_{\l(\pi_0\circ\pi_1\circ\pi_2\circ\ldots\circ \pi_p\rr)}.
\ee
We know the probability of each eventuality exactly, as explained in
Sec. (\ref{history}). The remaining part is the configurational entropy. As
explained above, we can calculate this entropy analytically for
simple configurations. But to obtain a conclusive answer, we carry out
a higher level calculation using the computer program
Mathematica. Using the reduction trick of Sec. \ref{srr}, we can simplify
any configuration of level-n to a density matrix involving at most
$4+3n$ spins. This allows an exact numerical computation. 

We carried out this computation up to level-3 configurations. The
result for the average configurational entropy was:
\be
\S_{total}=1.3327 - 0.001.
\ee
Note that this answer is an overestimate (hence the subtraction sign
for error indication), and thus can not be taken to
be exactly 4/3. The closeness of the answer to 4/3, however, is quite
mysterious. 

\subsection{Effective central charge results}

The effective central charge of the critical spin-1 chain is thus:
\be
c_{eff}^{(r_c)}=\frac{4}{3}\cdot 1.3327\cdot \ln 2 = 1.232
\ee
and the entanglement entropy  between a segment of length L and the
rest of the chain is:
\be
\frac{1}{3} 1.232 \log_2 L.
\label{answer}
\ee
where we use the level-3 result above. Note that $c_{eff}^{(r_c)}= 16/9-\epsilon$
where $\epsilon\approx 10^{-4}$. In the next section we discuss the
significance of our results in the context of infinite-randomness
fixed points.

\section{Discussion \label{discussion}}

Several random quantum critical points in 1D are now known to have
logarithmic divergences of entanglement entropy with universal
coefficients, as in the pure case.  So far, all analyzed systems were infinite randomness critical points in the
Random-singlet universality class.
\cite{RefaelMoore2004,Santachiara,Laflorencie,
  YangBonesteel} Summarizing the results for these systems is
possible with the formula:
\be
S\sim\frac{1}{3}\ln D \log_2 L,
\ee
where $D$ is the dimension of the gapless sector of the Hilbert-space
on each site, which generalizes to the quantum dimension in the case
of the golden-chain. \cite{YangBonesteel} Hence the 'effective central
charge' for these critical points is simply:
\be
c_{eff}=\ln D.
\label{RSsimple}
\ee

The present work investigates the disorder-averaged entanglement
entropy of a different universality class within the infinite-randomness
framework: the spin-1 Haldane-RS critical point. We indeed find a
different result than Eq. (\ref{RSsimple}), which arises from a more
complex structure of the spin-1 Haldane-RS point. We find the
entanglement entropy for this fixed point to be:
\be
S\sim\frac{1}{3}{c_{eff}}\log_2 L=\frac{1}{3}\frac{4}{3}\cdot
1.3327\cdot \ln 2 \log_2 L.
\ee
Our calculation required both the exact probabilistic description of the
low-energy structure given by RSRG, and the individual determination
of the quantum entanglement of each spin configuration in the
low-energy structure.

An open question is whether other physical properties are
related to the universal coefficient.  In pure 1D quantum critical
systems described by 2D classical theories with conformal invariance,
the central charge $c$ controls several important physical properties
beyond the entanglement entropy.  An example is the specific heat at
low temperatures: because the central charge determines the change in
free energy when the 2D classical system is compactified in one
direction (i.e., put on a cylinder), the quantum system at finite
temperature has a contribution to the specific heat that is
proportional to $c$.

The central charge also gives some information about the
renormalization-group flows connecting different critical points
because of Zamolodchikov's $c$-theorem: \cite{Zamolodchikov} RG flows
are from high central charge to low central charge, compatible with
the heuristic view of the RG as ``integrating out'' degrees of
freedom.  Ref. \onlinecite{FradkinCNeto} goes as far as defining a
c-function for a finite-temperature quantum system, which decreases with
decreasing temperatures, as it flows to a stable conformally-invariant
fixed point. It is logical to ask~\cite{Vidal03,RefaelMoore2004} whether
entanglement entropy can similarly be used to predict the direction of
real-space RG flows. As pointed out in the introduction, Sec. \ref{intro}, this
question separates in two: 
\begin{enumerate}
\item \label{ia} flows from pure conformally-invariant
points to infinite randomness points, 
\item \label{ib} flow between two
different infinite-randomness fixed points.  
\end{enumerate}

Recently, a
counterexample for question (\ref{ia}) was found by
Santachiara:\cite{Santachiara} for the random quantum parafermionic
Potts model with $N>41$, the random critical point was found to have larger entanglement than
the pure critical point, even though randomness is relevant at the
pure critical point. This point too obeys Eq. (\ref{RSsimple}) with
$D\rightarrow N$. Until now, there were no models in which we could
ask the second question; the spin-1 random Heisenberg model provides
the first test of case (\ref{ib}).

The critical point of the spin-1 Heisenberg model we found the effective central charge
\be
c^{(r_c)}_{eff}=1.232\approx \frac{16}{9}\ln 2.
\ee
As shown in Fig. \ref{spin1PD}, this point is unstable towards the spin-1 RS point with entanglement
entropy:
\be
c^{RS}_{eff}=\ln 3=1.099<c^{(r_c)}_{eff}.
\ee
On the weak randomness side, the critical point is unstable towards
the Haldane phase (with topological order, but with a suppressed gap
due to Griffiths effects. \cite{Damle-griffiths,mdhGriffith} 
The Haldane phase does not have a $\ln L$ term at
all. Hence, within the critical points of the random Heisenberg model, effective $c$ does
decrease along flows, which agrees with case \ref{ib} above. 

The Haldane-RS critical point of the random spin-1 chain is
thought to be the terminus of a flow beginning with the pure spin-1
chain in Eq. (\ref{TB}), which is a $k=2$ WZW theory with central
charge $c=3/2$. Since $c^{(r_c)}_{eff}=1.232<3/2$, the effective
Central charge indeed also decreases along the flow line from the pure
to the random critical fixed point, as posited in case \ref{ia}
above. 

In order to make clear the similarity between the random-singlet phase
of the spin-1 chain and the random-singlet phase of the spin-1/2
chain, it seems worthwhile to introduce a modified central charge that
is the coefficient of the $1/3 \log_2 L$ divided by the measure of the
local ungapped Hilbert space in each site, $\ln D$. This clearly puts
all random singlet phases on the same footing even when arising in
different microscopic systems.  In addition, one may ask whether such
a redefined measure may always reduce along RG flows, even those
connecting pure fixed points to random ones. 

An interesting difference between the spin-1 case we study and
 previous work, is that for RS phases the exact logarithmic divergence of entanglement
entropy can be found in closed form using the RG history approach,
while for spin-1, the true entanglement entropy seems to depend on the
entanglement values of an infinite number of nonequivalent, irreducible
sub-chains. This turn of events motivated us to define the 'reduced
 entropy' in Sec. \ref{reduced}. This simplified measure of the
 entanglement counts directly how many SBS's connect the segment $L$
 with the rest of the chain, and neglects the correlation induced by
 the SBS's. Our analysis allowed an exact calculation of the reduced
 entanglement, which yielded:
\be
c^{reduced}_{eff}=\frac{23}{12}\ln 2\approx 1.329.
\ee
A numerical test using a computerized application of the RSRG
 confirmed this exact analytical result (see Sec. \ref{numerics}),
 thus affirming the history-dependence segment of our approach.  

In light of the simplicity of the definition of the reduced
entanglement, it is interesting to ask how is relates to the real
quantum entanglement. Since it neglects correlations between singlets,
the reduced entanglement is most likely an {\it overestimate} of the
quantum entanglement. This remains to be proved generally. In the
spin-1/2 case, the reduced entanglement and the real entanglement
coincide. For the spin-1 RS-Haldane critical point, we have:
\be
c^{reduced}_{eff}\approx 1.329>1.232\approx c^{(r_c)}_{eff}
\ee
Hence the reduced entropy is bigger by about 10$\%$.  If there were some simplifying relation in the
entanglement of these irreducible units, as there is for the reduced
entanglement, then our method would give a closed form, but barring
that it seems that entanglement entropy is a less natural object in
the real-space renormalization group for higher spin than the (unphysical)
reduced entropy.

Future directions include, needless to say, the calculation of the
universal entanglement entropy in other infinite-randomness non RS
fixed points. Our method can be straightforwardly applied to $s>1$
chains. But an exciting possibility is the consideration of the
infinite-randomness fixed points arising in non-abelian
spin chains.\cite{RMYB}  The random-singlet subset of this class
of fixed points were recently studied in Ref. \onlinecite{YangBonesteel}.

\acknowledgments

We gratefully acknowledge useful conversations with A. Kitaev, I. Klich,
A. W. W. Ludwig, and J. Preskill, R. Santachiara, the hospitality of the Kavli Institute for Theoretical Physics, and support from NSF grants PHY99-07949 and DMR-0238760.

\vspace{5mm}

\appendix

\section{Calculation of
  $N_{\downarrow\downarrow},\,N_{\uparrow\downarrow}$ \label{appA}}

In Sec. \ref{srr} we defined the states:
\be
\ba{c}
\ket{\uparrow~(1,1)~\uparrow}=
\ad{0}\ad{n}\prodd_{i=0}^{n-1}\l(\ad{i}\bd{i+1}-\bd{i}\ad{i+1}\rr)\ket{0}\vspace{2mm}\\
\ket{\uparrow~(1,1)~\downarrow}=
\ad{0}\bd{n}\prodd_{i=0}^{n-1}\l(\ad{i}\bd{i+1}-\bd{i}\ad{i+1}\rr)\ket{0}\vspace{2mm}\\
\ket{\downarrow~(1,1)~\uparrow}=
\bd{0}\ad{n}\prodd_{i=0}^{n-1}\l(\ad{i}\bd{i+1}-\bd{i}\ad{i+1}\rr)\ket{0}\vspace{2mm}\\
\ket{\downarrow~(1,1)~\downarrow}=
\bd{0}\bd{n}\prodd_{i=0}^{n-1}\l(\ad{i}\bd{i+1}-\bd{i}\ad{i+1}\rr)\ket{0}.
\ea
\ee
We found that these states do not constitute a orthogonal set since:
\be
\langle
\uparrow~(1,1)~\downarrow\ket{\downarrow~(1,1)~\uparrow}=(-1)^n.
\ee

To be able to carry out the procedure outlined in Sec. \ref{srr} we
need to calculate the norms of these states, which will than allow a
transformation into an orthonormal combination. 

From standard transfer matrix techniques, and Wick contractions, we
find for $N_{\uparrow\uparrow}$:
\begin{widetext}
\be
\ba{c}
N_{\uparrow\uparrow}=\bra{0}\l(\a{0}\a{n}\prodd_{i=0}^{n-1}\l(\a{i}\b{i+1}-\b{i}\a{i+1}\rr)\rr)\ad{0}\ad{n}\prodd_{i=0}^{n-1}\l(\ad{i}\bd{i+1}-\bd{i}\ad{i+1}\rr)\ket{0}\vspace{2mm}\\
=\l(\ba{c} 2 \vspace{2mm}\\ 1 \ea \rr)^T\l(\ba{cc} 1 & 2 \vspace{2mm}\\ 2 & 1 \ea \rr)^{n-1} \l(\ba{c} 1 \vspace{2mm}\\ 2 \ea \rr)\vspace{2mm}\\
=\l(\ba{c} 2 \vspace{2mm}\\ 1 \ea \rr)^T\frac{1}{\sqrt{2}}\l(\ba{cc} 1 & 1 \vspace{2mm}\\ 1 & -1 \ea \rr) 
\l(\ba{cc} 3 & 0 \vspace{2mm}\\ 0 & -1 \ea \rr)^{n-1} \frac{1}{\sqrt{2}}\l(\ba{cc} 1 & 1 \vspace{2mm}\\ 1 & -1 \ea \rr)\l(\ba{c} 1 \vspace{2mm}\\ 2 \ea \rr)\vspace{2mm}\\
=\frac{1}{2}\l(\ba{c} 3 \vspace{2mm}\\ 1 \ea \rr)^T\l(\ba{cc} 3 & 0 \vspace{2mm}\\ 0 & -1 \ea
\rr)^{n-1}\l(\ba{c} 3 \vspace{2mm}\\ -1 \ea \rr)=\frac{1}{2}\l(3^{n+1}+(-1)^n\rr).
\ea
\ee

Very similarly, for $N_{\uparrow\downarrow}$:
\be
\ba{c}
N_{\uparrow\downarrow}=\bra{0}\l(\a{0}\b{n}\prodd_{i=0}^{n-1}\l(\a{i}\b{i+1}-\b{i}\a{i+1}\rr)\rr)\ad{0}\bd{n}\prodd_{i=0}^{n-1}\l(\ad{i}\bd{i+1}-\bd{i}\ad{i+1}\rr)\ket{0}\vspace{2mm}\\
=\l(\ba{c} 2 \vspace{2mm}\\ 1 \ea \rr)^T\l(\ba{cc} 1 & 2 \vspace{2mm}\\ 2 & 1 \ea \rr)^{n-1} \l(\ba{c} 2 \vspace{2mm}\\ 1 \ea \rr)\vspace{2mm}\\
=\l(\ba{c} 2 \vspace{2mm}\\ 1 \ea \rr)^T\frac{1}{\sqrt{2}}\l(\ba{cc} 1 & 1 \vspace{2mm}\\ 1 & -1 \ea \rr) 
\l(\ba{cc} 3 & 0 \vspace{2mm}\\ 0 & -1 \ea \rr)^{n-1} \frac{1}{\sqrt{2}}\l(\ba{cc} 1 & 1 \vspace{2mm}\\ 1 & -1 \ea \rr)\l(\ba{c} 2 \vspace{2mm}\\ 1 \ea \rr)\vspace{2mm}\\
=\frac{1}{2}\l(\ba{c} 3 \vspace{2mm}\\ 1 \ea \rr)^T\l(\ba{cc} 3 & 0 \vspace{2mm}\\ 0 & -1 \ea
\rr)^{n-1}\l(\ba{c} 3 \vspace{2mm}\\ 1 \ea \rr)=\frac{1}{2}\l(3^{n+1}-(-1)^n\rr).
\ea
\ee
\end{widetext}

%\bibliography{bigbib}
\bibliography{entspinrefs}

\end{document}